\documentclass[twocolumn]{aastex6}
\usepackage{tikz}

\usepackage{amsmath}
\usepackage{amssymb}
\usepackage{epsfig}
\usepackage{graphicx}
\usepackage{ifthen}
\usepackage{latexsym}
\usepackage{rotating}
\usepackage{subfigure}
\usepackage{times,epsf}
\usepackage{txfonts}
\usepackage{varioref}
\usepackage{verbatim}
\usepackage{array}
\usepackage{url}
\usepackage{color}
\usepackage{multirow}

\newcommand{\be}{\begin{equation}}
\newcommand{\ee}{\end{equation}}
\newcommand{\bea}{\begin{eqnarray}}
\newcommand{\eea}{\end{eqnarray}}

\newcommand{\avg}[1]{\langle #1\rangle}

\newcommand{\koral}{\texttt{KORAL}\,}
\newcommand{\Medd}{\dot M_{\rm Edd}}

\shorttitle{Acceleration of wind in BH accretion disks}
\shortauthors{A. M\"{o}ller,~A. S\c{a}dowski}

\begin{document}

\title{Acceleration of
  wind in optically thin and thick black hole accretion disks simulated in general relativity}

\author{Anton M\"{o}ller}
\affil{KTH Royal Institute of Technology, Brinellv{\"a}gen 8, SE-11428 Stockholm, Sweden\\
Research Science Institute, MIT, 
77 Massachusetts Ave, Cambridge, MA 02139, USA}
\email{amoller@mit.edu}

\author{Aleksander S\c{a}dowski}
\affil{MIT Kavli Institute for Astrophysics and Space Research,
77 Massachusetts Ave, Cambridge, MA 02139, USA\\
Einstein Fellow}
\email{asadowsk@mit.edu}

\label{firstpage}

\begin{abstract}We study the force balance and resulting acceleration
  of gas in general relativity basing on simulations of accretion on a
  stellar-mass, non-rotating black hole. We compare properties of
  acceleration in an optically thin, radiatively inefficient disk, and
  in an optically thick, super-critical disk accreting at 10 times the
  Eddington rate. We study both the average forces acting at given
  location and forces acting on a gas parcel along its 
  trajectory. We show that the acceleration is not a continuous
  process -- in most cases gas is accelerated only in short-lasting
  episodes. We find that in the case of optically thin disks gas is
  pushed out by magnetic field in the polar region and by thermal
  pressure and centrifugal force below the disk surface. In case of
  optically thick, radiative accretion, it is the radiation pressure
  which accelerates the gas in the polar funnel and which compensates
  (together with the centrifugal force) the gravity in the bulk of the
  disk. We also show that the Newtonian formulae for the forces are
  inadequate in the innermost and in the highly magnetized regions.
\end{abstract}

\keywords{accretion, accretion discs -- black hole physics -- relativistic
  processes -- methods: numerical -- galaxies: jets}

\section{Introduction}\label{s.intro}

Gas accreting on compact objects liberates large amounts of its
binding energy. Even small fraction of this energy converted into
radiation makes accreting systems extremely luminous. But radiation is
not the only product of accretion. There is a growing evidence that
accretion inevitably generates outflows carrying outward significant amount of
gas -- sometimes much more than what reaches the compact object. 

Gas may be driven out of the accretion flows in a couple of
ways. Large scale magnetic fields may lead to magneto-centrifugal
acceleration of gas along rapidly rotating field lines
\citep{blandfordpayne, sadowskisikora-10}. Radiation generated by the
infalling gas may exert outward  pressure that accelerates the
gas \citep{ss73}. Finally, illumination of the outer regions by hot radiation
coming from the inner region may drive thermal outflow \citep{begelman+83}.

Blueshifted absorption features indicating the presence of outflowing
gas along the line of sight have been observed for years both in case of
accreting stellar mass compact objects and active galactic nuclei (AGN)
\citep{kingashley+13}. It has been shown that winds are most prevalent
in the soft state of black hole (BH) X-ray binaries \citep{ponti+12}.
Their presence is anticorrelated with the presence of radio
jets \citep{miller+08,neilsen+09}. They are highly variable
\citep[e.g.,][]{neilsen+11,neilsen+12,neilsen-13} and carry matter out at significant
rates \citep[e.g.,][]{ueda+04,kallman+09,miller+16}.

Deriving the properties of the outflow from the observed absorption
lines is not straightforward. The observed absorption occurs along the
line of sight as a
result of interplay between gas density and ionization, often
far from the region where the outflow was generated. In addition, the absorption
profiles are superimposed on the relativistically broadened reemission
features coming from larger volume \citep{miller+15}.
There is no
consensus on whether the observed outflows in X-ray binaries are
driven magneto-centrifugally
\citep{miller+06,miller+08,kallman+09,neilsen+16,miller+16}, thermally \citep{ueda+09,neilsen+11,neilsen+14,diaztrigo+14},
or whether both mechanisms work at the same time
\citep{neilsen+12}. Active galactic nuclei show significant
outflows as well \citep{king+11,kingashley+12}, some of which can be
mildly relativistic \citep{king+15,chartas+02,tombesi+ufos}, providing
physical mechanism for connecting the SMBH and the properties of its host.

Analytical modeling of accretion flows usually takes strong
assumptions, e.g., by neglecting the outflows completely
\citep{narayanyi-94} or
prescribing their radial dependence \citep{adios}, and does not allow to study their
generation consistently. Multi-dimensional simulations are necessary
for this purpose. However, they are limited, and a single simulation
is not able to cover the whole accretion flow. \cite{woods+96}, \cite{luketic+10} and \cite{higginbottom+15} focused on the outer regions and studied 
the
generation of thermal winds from the photosphere irradiated by strong
X-rays produced in the innermost part. Other groups
\citep[e.g.,][]{proga+00,gammie03,ohsuga09,skadowski2014numerical}
studied the innermost region where magnetocentrifugal and radiative
driving operates and showed that, indeed, singificant outflows emerge.

In this paper we study the acceleration mechanisms acting on outflowing
gas in optically thin and optically thick accretion flows 
by analyzing simulations of the innermost region performed with general relativistic radiation magnetohydrodynamical (GRRMHD) code \koral.
To model these phenomena, we break down the individual forces acting
upon the gas and determine the main acceleration mechanisms in each case.
 
Similar modelling has been done previously by~\cite{yuan2015numerical}
and \cite{takahashi2015radiation}, but this paper
presents the first general relativistic modeling and directly compares
driving mechanisms in optically thin and thick disks performed within
the same computational framework. We study only accretion flows on non-rotating, stellar mass BHs, and limit ourselves to accretion with weak, not-saturated magnetic fields \citep[SANE, in the formalism put forward in][]{narayan2012grmhd}.

The paper is organized as follows. In Section~\ref{s.forces}, we
decompose the forces acting on a particle. Section~\ref{s.method}
describes the simulation environment and the initial
conditions. Sections~\ref{s.results} presents the results and
Section~\ref{s.discussion} discusses them and gives concluding remarks.

\subsection{Definitions}\label{s.char.quant}

We adopt the following definition of the Eddington accretion rate,
\begin{equation}
	\dot{M}_\mathrm{Edd} = \frac{L_\mathrm{Edd}}{\eta_0 c^2},
\end{equation}
where 
\begin{align}
	\eta_0 &= 1-\sqrt{1-\frac{2}{3R_\mathrm{ISCO}}},
\end{align}
is the efficiency of a thin disk \citep{ss73,nt73} extending down to
the marginally stable orbit located at $R_\mathrm{ISCO}$. According to
this definition, a thin disk accreting at $\dot{M}_\mathrm{Edd}$ emits
the Eddington luminosity. For zero BH spin,
$\Medd = 2.48 \times 10^{18}M/M_{\odot}  \,\rm g/s$.

We use the term \textit{outflow} to denote any consistent motion with
positive radial velocity $v_r$. This definition includes both the \textit{real
  outflow} and the \textit{turbulent outflow}. While the former
eventually reaches infinity, the latter eventually rejoins the
accretion flow due to the turbulent gas motion in the disk To
determine whether or not a flow is part of the real outflow, we define
the relativistic Bernoulli parameter $Be$ \citep{sadowski+3d},
\begin{equation}
\label{eq:Be}
	Be = -\frac{T^t_t+R^t_t+\rho u^t}{\rho u^t},
\end{equation}
where $T$, $R$, $\rho$ and $u^\mu$ are gas and radiation stress-energy
tensors, gas density and four-velocity, respectively. The real outflow
will satisfy $Be>0$. However, it is important to note that the
Bernoulli parameter of a flow will often fluctuate. Despite having a
negative parameter when it reaches the simulation boundaries, the flow
might still be part of the real outflow, if it ultimately reaches
$Be\ge 0$. This can typically be seen as a steady increase in the
parameter over the course of the trajectory~\citep{yuan2015numerical}.

We define \textit{polar funnel} as the conical region of low density between the disk surface
and the polar axis.

Hereafter, we use the
gravitational radius $r_{\rm g}=GM_{\rm BH}/c^2$ as the unit of length, and $r_g/c$
as the unit of time, and often assume $c=1$. We also always assume that the underlying spacetime metric is fixed, i.e., is not perturbed by the accretion flow.

\section{Force decomposition}\label{s.forces}

\subsection{Conservation laws}
The three general conservation laws, that is conservation of mass, energy, and angular momentum, are described by the following set of equations,
\begin{align}
\label{e.cons_mass}	(\rho u^\mu)_{;\mu} &= 0, \\
\label{e.cons_T}	(T^\mu_\nu)_{;\mu} &= G_\nu, \\
\label{e.cons_R}	(R^\mu_\nu)_{;\mu} &= -G_\nu,
\end{align}
where $\rho$ is the gas density in the comoving frame, $u^\mu = \left( u^t ,u^i \right)$ is the angular four-velocity, semicolon indicates the covariant derivative, and $T^\mu_\nu$ is the magnetohydrodynamic stress-energy tensor,
\begin{equation}\label{e.stress-energy-tensor}
	T^\mu_\nu =wu^\mu u_\nu + \left( p_{\mathrm{g}}+\frac{1}{2}b^2\right)\delta^\mu_\nu-b^\mu b_\nu.
\end{equation}
$w = \rho+u_\mathrm{g}+p_\mathrm{g}+b^2$ denotes relativistic
enthalpy, $u_\mathrm{g}$ and $p_\mathrm{g}=\left( \gamma -1
\right)u_\mathrm{g}$ are the internal energy and gas pressure, respectively, $\gamma=5/3$ is the adiabatic
index, $b^\mu$ is the magnetic field four-vector and
$\delta^\mu_\nu$ is the Kronecker delta.

$R^\mu_\nu$ is the stress-energy tensor of radiation and $G_\nu$ is the radiation four-force describing the interaction between gas and radiation defined by,
\begin{equation}\label{e.rad_four}
	G^\mu = G^\mu_{0} + G^\mu_{\mathrm{Compt}},
\end{equation}
where $G^\mu_{0}$ and $G^\mu_{\mathrm{Compt}}$ reflect the energy and momentum transfer due to absorption and scattering, and Comptonization, respectively. The former is given by \citep{skadowski2015global},
\begin{equation}\label{e.rad_four_0}
G^\mu_{0} = -\rho(\kappa_a + \kappa_s)R^{\mu\nu}u_\nu-\rho\left( \kappa_s R^{\alpha\beta}u_\alpha u_\beta + \kappa_a 4\pi B \right) u^\mu,
\end{equation}
where $\kappa_a$ and $\kappa_s$ are the absorption and scattering grey
opacities respectively and  $4\pi\widehat{B} = aT_{\rm g}^4$  ($a$ is the radiation constant) is the intensity of black body radiation for a gas with temperature $T_{\rm g}$. We further define an effective radiation temperature $T_{\rm r}$ in the fluid frame as
\begin{equation}
	\widehat{E} = aT_{\rm r}^4.
\end{equation}
The Comptonization component of Equation~\eqref{e.rad_four}, under the ``blackbody'' approximation \citep[see][]{sadowski+compt} is,
\begin{align}
\begin{split}
G^\mu_\mathrm{Compt} =& -\kappa_s\rho\widehat{E}\left[\frac{4k(T_{\rm
      g}-T_{\rm r})}{m_ec^2}\right]\times\left[ 1+3.683\left(
    \frac{kT_{\rm g}}{m_ec^2} \right) + 4\left( \frac{kT_{\rm g}}{m_ec^2} \right)^2 \right] \\ 
	&~~~~ \times \left[ 1+\frac{kT_{\rm g}}{m_ec^2} \right]^{-1}u^\mu,
\end{split}
\end{align}
where  $m_e$ is the electron mass. 

In the orthonormal fluid frame, equation~(\ref{e.rad_four_0}) corresponds to the more intuitive 
\begin{align}
	\label{e.rad_four_t}	\widehat{G}^t &= \kappa_a\rho\left(\widehat{E}-4\pi\widehat{B}\right), \\
	\label{e.rad_four_i}	\widehat{G}^i &= (\kappa_a+\kappa_s)\rho\widehat{F}^i,
\end{align}
where $\widehat{F}$ is the radiation flux. Equation~\eqref{e.rad_four_t} describes the energy transfer rate resulting from absorption and emission while Equation~\eqref{e.rad_four_i} describes the rate of change of the momentum because of absorptions and scatterings.

Equations~\eqref{e.cons_mass}-\eqref{e.cons_R} can be rewritten in a coordinate basis by separating the time dimension from the spatial coordinates yielding,
\begin{align}
\label{e.cons_coord_mass}	\partial_t (\rho u^t) + \partial_i (\sqrt{-g}\rho u^i) &= 0 \\
\label{e.cons_coord_T}	\partial_t (T^t_\nu) + \partial_i (\sqrt{-g}T^t_\nu) &= \sqrt{-g}T^\kappa_\lambda\Gamma^\lambda_{\nu\kappa} + \sqrt{-g}G_\nu \\
\label{e.cons_coord_R}	\partial_t (R^t_\nu) + \partial_i (\sqrt{-g}R^t_\nu) &= \sqrt{-g}R^\kappa_\lambda\Gamma^\lambda_{\nu\kappa} - \sqrt{-g}G_\nu
\end{align}
where $\sqrt{-g}$ is the metric determinant and $\Gamma^\lambda_{\nu\kappa}$ are the Christoffel symbols.

\subsection{Decomposition}

To decompose the forces acting on the gas, we rewrite Equation~\eqref{e.cons_coord_T} assuming a stationary state ($\partial_t=0$) and decomposing the stress-energy tensor $T^\mu_\nu$ from Equation~\eqref{e.stress-energy-tensor}, yielding,
\begin{align}
\begin{split}\label{e.T_decompose}
	\partial_i (\sqrt{-g}wu^iu_\nu) =& -\partial_i (\sqrt{-g}p_\mathrm{g}\delta^i_\nu ) -\partial_i \left(\sqrt{-g}\frac{1}{2}b^2\delta^i_\nu\right) \\ &~~~~+ \partial_i (\sqrt{-g}b^ib_\nu) + \sqrt{-g}T^\kappa_\lambda\Gamma^\lambda_{\nu\kappa} + \sqrt{-g}G_\nu.
\end{split}
\end{align}
The left hand side of Equation~\eqref{e.T_decompose} equals,
\begin{align}\label{e.T_decompose_LHS}
\begin{split}
	\partial_i (\sqrt{-g}wu^iu_\nu) &= u_\nu\cdot\partial_i (\sqrt{-g}wu^i)+ \sqrt{-g}wu^i\cdot\partial_i (u_\nu) \\
	&= u_\nu\cdot\partial_i (\sqrt{-g}(w-\rho) u^i)+ \sqrt{-g}wu^i\cdot\partial_i (u_\nu),
\end{split}
\end{align}
where $\partial_i (\sqrt{-g}\rho u^i)=0$ (eq.~\ref{e.cons_coord_mass}) due to the assumption of a stationary state. We move the metric determinant out of the derivatives and combine this with Equations~\eqref{e.T_decompose} and~\eqref{e.T_decompose_LHS} to obtain,
\begin{align}
\begin{split}
	u^i\partial_i (u_v) &= -\frac{1}{w}\partial_\nu p_\mathrm{g}-\frac{1}{w}\partial_\nu\left(\frac{1}{2}b^2\right) + \frac{1}{w}\partial_i (b^ib_\nu) + \frac{1}{w}T^\kappa_\lambda\Gamma^\lambda_{\nu\kappa} \\ 
	&-\frac{T^i_\nu-\rho u^iu_\nu}{w}\frac{\partial_i (\sqrt{-g})}{\sqrt{-g}}+\frac{1}{w}G_\nu - \frac{1}{w}u_\nu\partial_i\left( (w-\rho)u^i \right).
\end{split}
\end{align}
Here, the left hand side corresponds to the convective derivative of
the gas, while the terms on the right hand side correspond to
different forces acting on the gas. We define them as follows,
\begin{align}
\label{e.force_thermal}	f_{\nu,\mathrm{thermal}} &= -\frac{1}{w}\partial_\nu (p_\mathrm{g}) \\
\label{e.force_magentic}	f_{\nu,\mathrm{magnetic}} &= -\frac{1}{w}\partial_\nu \left( \frac{1}{2	}b^2\right) + \frac{1}{w}\partial_i (b^ib_\nu) \\
\label{e.force_metric}	f_{\nu,\mathrm{metric}} &= \frac{1}{w}T^\kappa_\lambda\Gamma^\lambda_{\nu\kappa} - \frac{T^i_\nu-\rho u^iu_\nu}{w}\frac{\partial_i (\sqrt{-g})}{\sqrt{-g}} \\
\label{e.force_radiation}	f_{\nu,\mathrm{radiation}} &= \frac{1}{w}G_\nu \\
\label{e.force_enthalpy}	f_{\nu,\mathrm{enthalpy}} &= -\frac{1}{w}u_\nu\partial_i \left((w-\rho)u^i\right).
\end{align}

Equation~\eqref{e.force_thermal} describes the thermal pressure
force. The thermal force is dependent on the gradient of the gas
pressure $\partial_\nu(p_\mathrm{g})$. This indicates that the force
is due to the internal pressure the gas exerts on itself (or rather,
the pressure the gas as a whole exerts on individual particles).

Equation~\eqref{e.force_magentic} is composed of two terms, the first
which shall be referred to as the magnetic pressure and the second as
the magnetic tension. Like the thermal force, the magnetic pressure is
a gradient force. This force arises from the density of the magnetic
field lines. Magnetic tension, on the other hand, is the force exerted
on particles as bent magnetic field lines straighten out.

Equation~\eqref{e.force_metric} describes the forces resulting from
the space time curvature. Most importantly, one of its components
reflects the gravitational acceleration. Another component, reflects
the (virtual from the point of view of a stationary observer)
centrifugal force. The remaining terms account mostly for the
relativistic corrections due to space-time curvature and are
negligible far from the BH.

Equation~\eqref{e.force_radiation} describes the force that radiation
exerts on particles. Photons can either accelerate particles due to
radiation pressure, or decelerated them because of the radiative drag,
as discussed above.

Equation~\eqref{e.force_enthalpy} is a relativistic correction
resulting from the difference between the total system enthalpy $w$
and the rest mass density $\rho$.

We can furthermore decompose Equation~\eqref{e.force_metric} into the components driving the gravitational attraction and the (azimuthal) centrifugal force,
\begin{align}
	f_{\nu,\mathrm{metric}} &= f_{\nu,\mathrm{gravity}} +f_{\nu,\mathrm{centrifugal}} + f_{\nu,\mathrm{residual}},
\end{align}
where,
\begin{align}
\label{e.force_gravity}	f_{\nu,\mathrm{gravity}} &= \frac{T^t_t}{w}\Gamma^t_{\nu t}\delta^\nu_r \\
\label{e.force_centrifugal} f_{\nu,\mathrm{centrifugal}} &= \frac{T^\phi_\phi}{w}\Gamma^\phi_{\nu \phi}.
\end{align}
The remainder, $f_{\nu,\mathrm{residual}}$, together with Equation~\eqref{e.force_enthalpy} reflects the relativistic correction terms, denoted by
\begin{equation}\label{e.force_correction}
	f_{\nu,\mathrm{correction}} = f_{\nu,\mathrm{residual}} + f_{\nu,\mathrm{enthalpy}}.
\end{equation}

The forces can be expressed in an orthonormal basis (assuming a diagonal metric of a non-rotating BH) through $\widehat{f_\nu} = f_\nu/\sqrt{g_{\nu\nu}}$.

For a non-relativistic system where $u_{\rm g}+p_\mathrm{g}+\frac{b^2}{2}\ll \rho$ and $v\ll 1$, Equations~\eqref{e.force_thermal}-\eqref{e.force_correction} reduce to the well known Newtonian formulae,
\begin{align}
\label{e.force_noGR_thermal}	f_{i,\mathrm{thermal}} &= -\frac{1}{\rho}\partial_i (p_\mathrm{g}) \\
\label{e.force_noGR_magentic}	f_{i,\mathrm{magnetic}} &= -\frac{1}{\rho}\partial_i \left( \frac{1}{2	}b^2\right) + \frac{1}{\rho}\partial_i (b^ib_i) \\
\label{e.force_noGR_metric}	f_{r,\mathrm{gravity}} &= -\frac{1}{r^2} \\
\label{e.force_noGR_centr_r}	f_{r,\mathrm{centrifugal}} &= \sin\theta\frac{v_\phi^2}{r\sin\theta} \\
\label{e.force_noGR_centr_th}	f_{\theta,\mathrm{centrifugal}} &= r\cos\theta\frac{v_\phi^2}{r\sin\theta} \\
\label{e.force_noGR_radiation}	f_{i,\mathrm{radiation}} &= \frac{1}{\rho}G_i\\
\label{e.force_noGR_correction}	f_{i,\mathrm{correction}} &= 0.
\end{align}

\section{Numerical Method}\label{s.method}

We analyzed two simulations performed with a general relativistic (GR)
radiation magnetohydrodynamical (RMHD) code
\texttt{KORAL}~\citep{sadowski+koral,skadowski2014numerical} - one
purely MHD (optically thin, \texttt{hd300a0}) and the other including radiation
field (optically thick, \texttt{d300a0}) and corresponding to the
accretion rate $\sim 10\dot M_{\rm Edd}$. Both were initialized as equilibrium
torii threaded with multiple loops of weak magnetic field lines. In
the case of the radiative run, local thermal equilibrium between gas
and radiation was assumed. The radiation was evolved adopting the M1
closure scheme. For both runs the BH spin was zero.

The simulations assumed axisymmetry and ran in $2.5$ dimensions, that
is, assuming axisymmetry but allowing non-zero azimuthal components.
We prevented quick dissipation of the magnetic field in
axisymmetrical simulations by applying the mean-field dynamo of
\cite{skadowski2015global}.

The simulations were run for an exceptional amount of time, reaching
$\sim 2,000$ orbits at the innermost stable circular orbit ($\sim
200,000\,GM_{\rm BH}/c^3$). This
allowed for the outflow to reach inflow/outflow equilibrium (defined
as the
region where average properties do not chang with time) in
relatively large domain, especially in the polar region, where the gas
moves with large velocities. Applying the criterion from
\cite{narayan2012grmhd} we obtain in both simulations the equilibirium regions extending up
to $\sim 80 r_{\rm g}$ at the equatorial plane and up to the computational
box boundary at $\sim 1000 r_{\rm g}$ at the axis.

For this work we adopted the following forms of the absorption and scattering opacities,
\begin{align}
 \kappa_a =& 6.4 \cdot 10^{22}~\rho T_{\rm g}^{-7/2}~\mathrm{cm}^2/g \\
 \kappa_s =& 0.34 ~\mathrm{cm}^2/g,
\end{align}
where $\rho$ and $T_{\rm g}$ are the gas density and temperature,
respectively.
Such a form of absorption opacities reflects only the bremsstrahlung
absorption and emission, and, in particular, neglects line
opacities. Therefore, the radiative simulation applies only to hot
accretion flows near stellar
mass BHs.

\section{Results}\label{s.results}

\subsection{Flow properties}

\begin{figure*}
	\centering
gas density and velocity (\texttt{hd300a0})\hspace{1.25cm}gas density and velocity (\texttt{d300a0})\hspace{1.5cm}radiation field (\texttt{d300a0})\hspace{2cm}
\rotatebox{90}{\hspace{3.4cm}{\large averaged}}\includegraphics[width=0.27\textwidth]{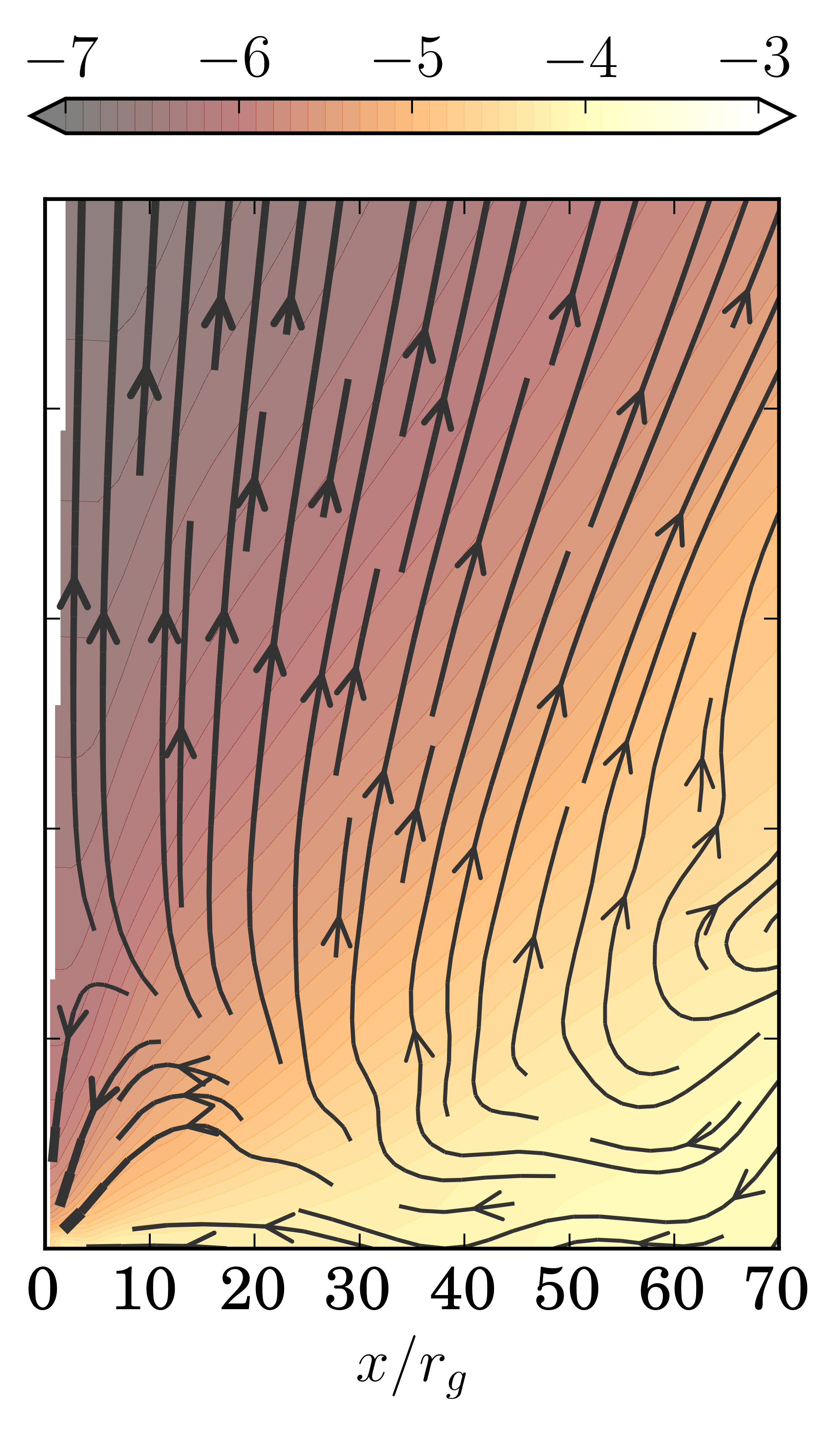}\hspace{.5cm}
\includegraphics[width=0.275\textwidth]{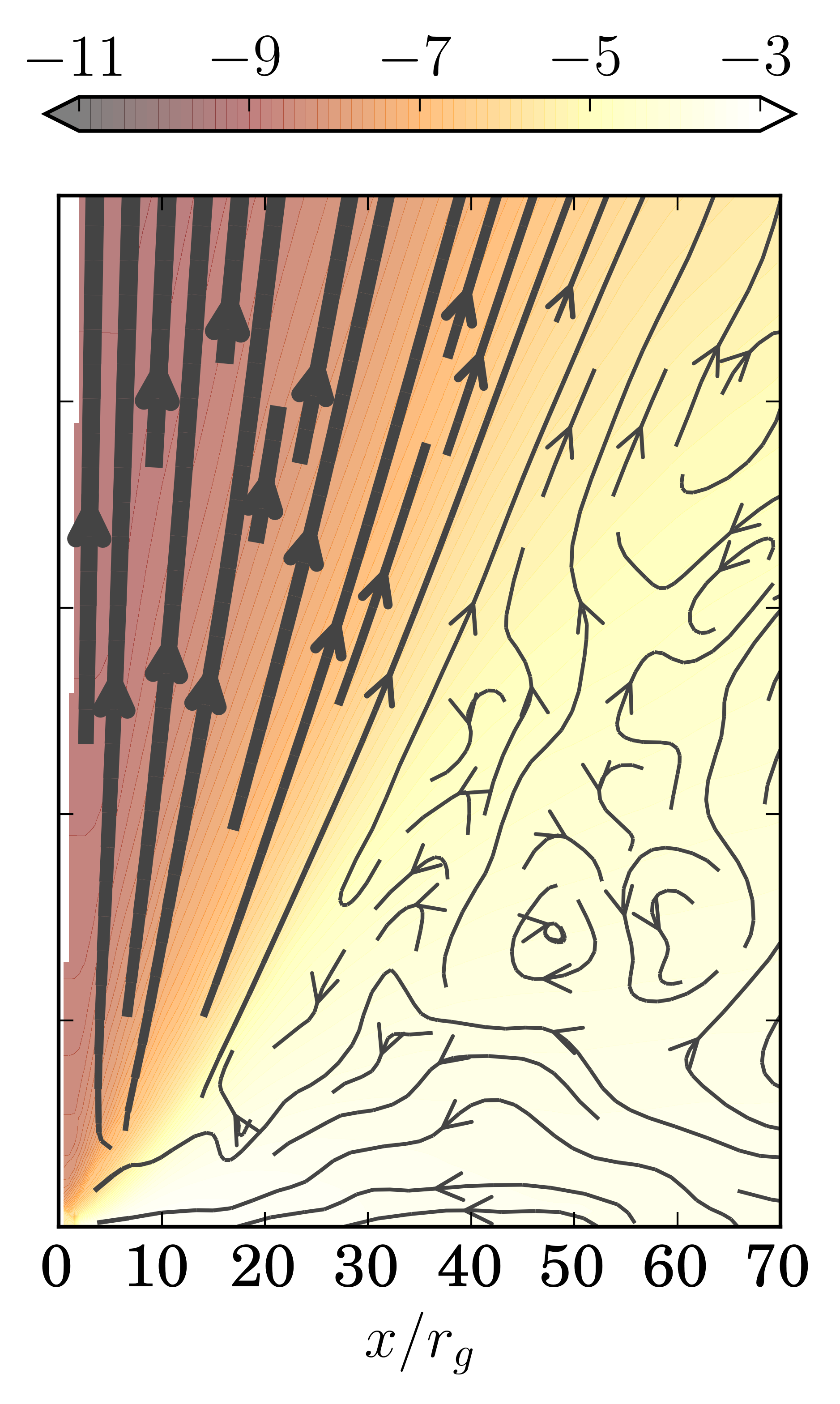}
\includegraphics[width=0.27\textwidth]{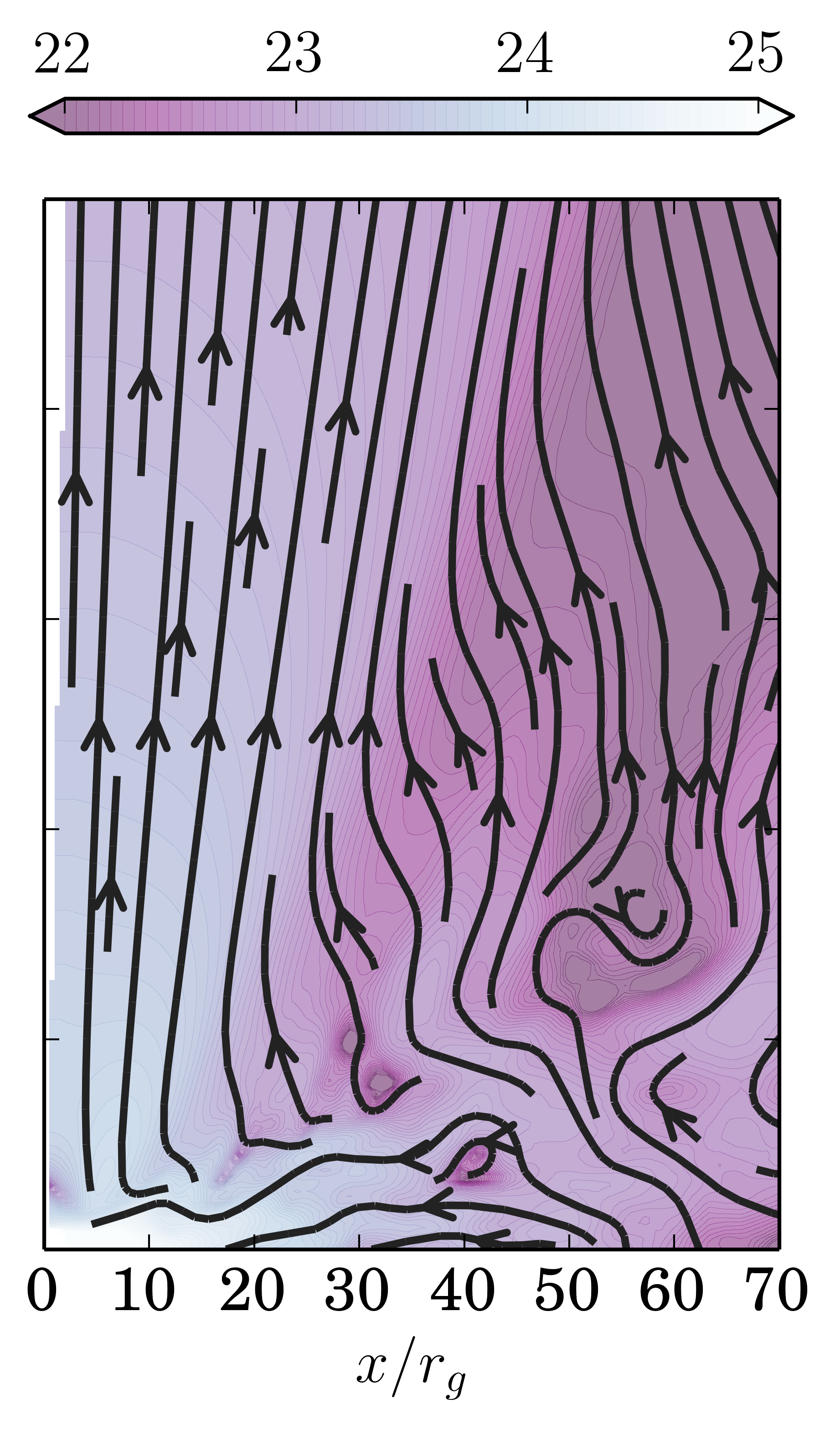}\vspace{-.65cm}
\rotatebox{90}{\hspace{3.4cm}{\large snapshot}}\includegraphics[width=0.27\textwidth]{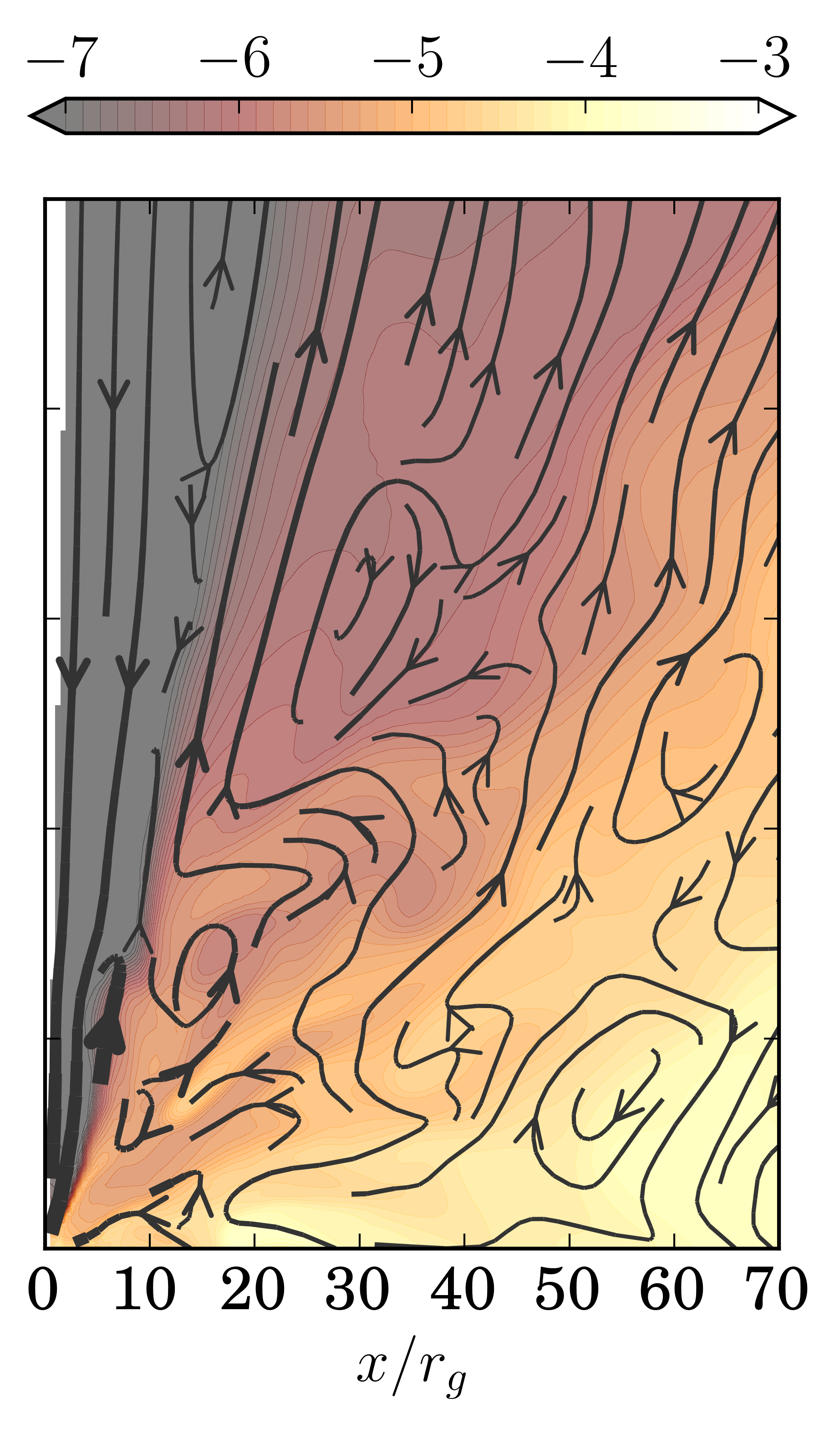}\hspace{.5cm}
\includegraphics[width=0.275\textwidth]{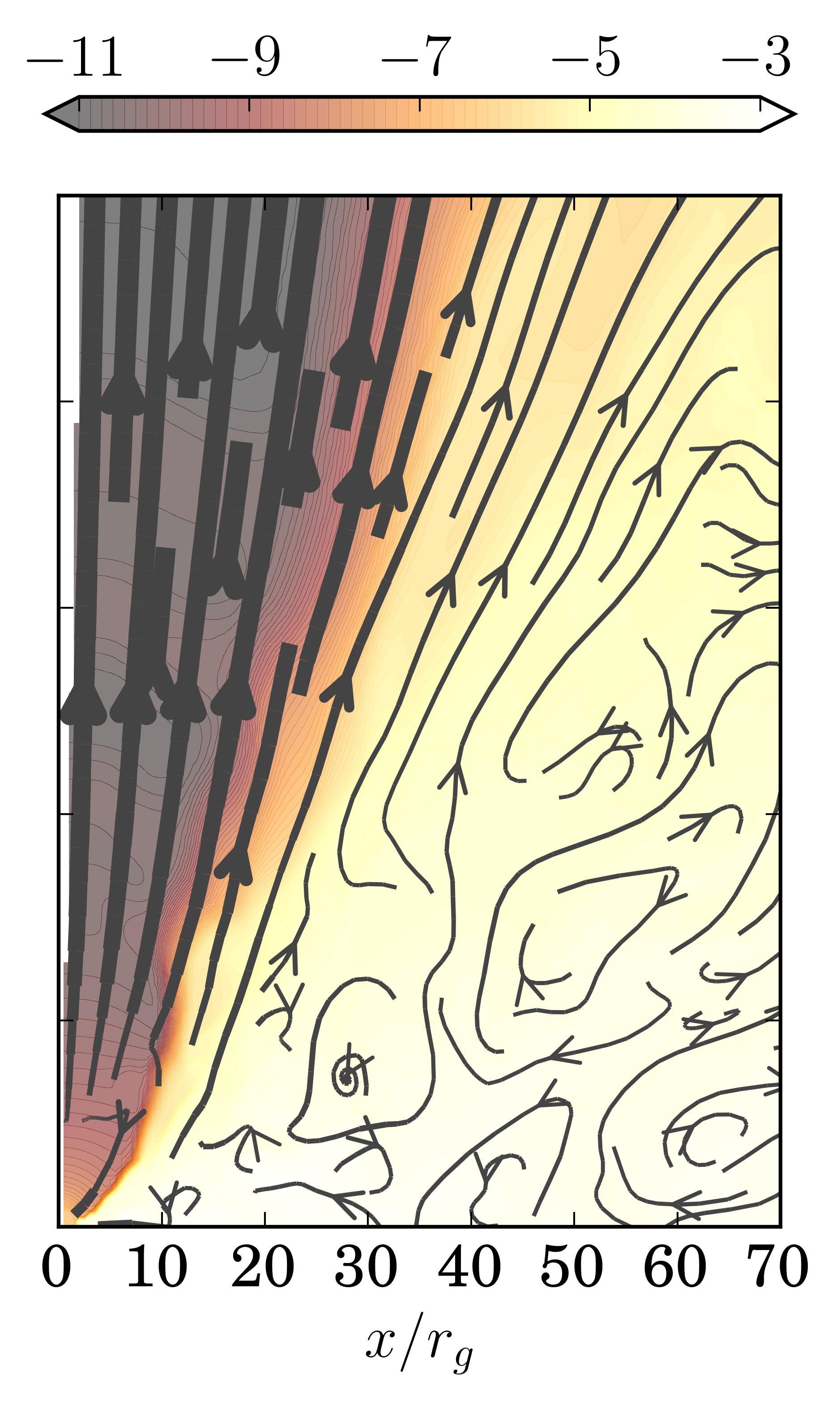}
\includegraphics[width=0.27\textwidth]{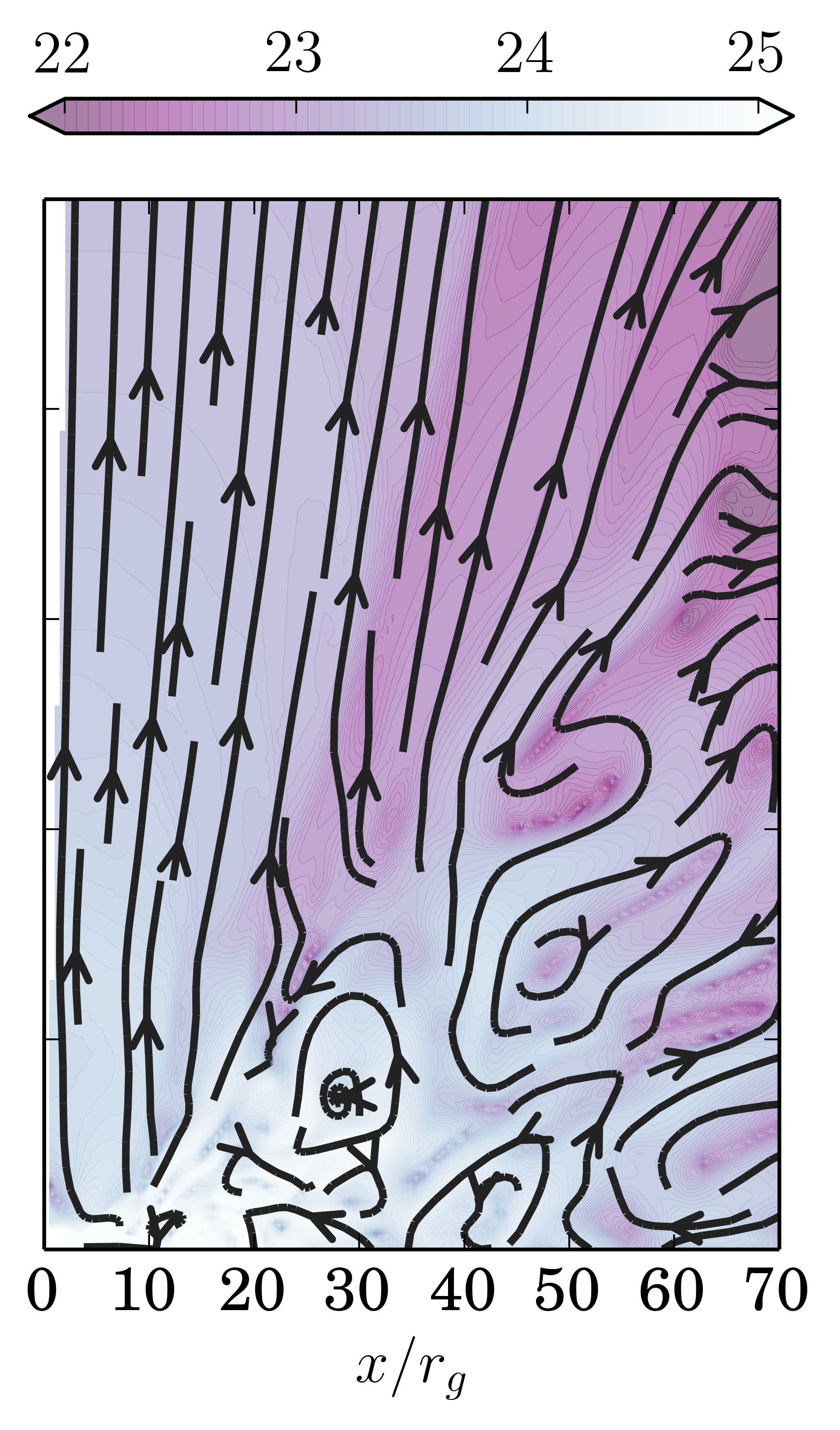}	
	\caption{Leftmost and center panels show the logarithm of the
          gas density distribution (in CGS units, $g/cm^3$) and the
          gas velocity field for the non-radiative and radiative
          simulations respectively. The width of the vectors are
          proportional to the velocity magnitude. The rightmost panels
          shows the logarithm of the magnitude of the radiative flux
          ($erg/s/cm^2$) and the radiative flux vector field. The top
          panels correspond to averaged data. The bottom panels
          represent instantaneous snapshot data at an arbitrary chosen
          time in the simulation.}
\label{f.snapshots.velocity.radflux}
\end{figure*}

The leftmost and center panels in
Figure~\ref{f.snapshots.velocity.radflux} show the averaged (top set
of panels) and instantaneous (bottom set) density distribution and
velocity vectors of gas in the non-radiative\footnote{Purely
  magnetohydrodynamical (non-radiative) simulations are scale-free, i.e., they correspond to
  optically thin flow of \textit{any} density. Therefore, the values
  of density given above the left panels are arbitrary.} and radiative
simulation respectively. The gas is clearly concentrated near the
equatorial plane but the disk thickness is large -- the density
scaleheight in both cases is close to $H/R=0.3$. In case of the hydro
simulation, the gas relatively smoothly fills the polar region. For
the radiative simulation, on the other hand, the polar region has
significantly lower density than the bulk of the disk. In that case,
the density at the axis can be as much as 6 orders of magnitudes lower
than at the equatorial plane. This fact results from radiation
cleaning the funnel, as will be
discussed below.

The streamlines in the left and center panels show the average
velocity field. In the case of the hydro simulation the gas flows
inward only deep inside the bulk of the disk. Far from the equatorial
plane vertical motion is clear, leading to gas escaping from the
disk. The velocity field in case of the radiative simulation is
noticeably different. Very fast (much faster than in the hydro case)
outflow is evident only in the polar region. At intermediate polar
angles, the gas does not show clear outflowing pattern, what implies
that the turbulent motion is strong and on average does not give
strong outflow.

The corresponding bottom panels show instantaneous gas density
distribution and velocity field. The turbulent nature of the flow,
resulting from the magnetorotational instability \citep{balbus1991powerful}, is evident in both cases. It is interesting to note that in the hydro simulation gas can temporarily fall on the BH, although on average it shows positive radial velocity. These episodes feature very low density gas which does not provide significant inward mass flux.

The right panels in Figure~\ref{f.snapshots.velocity.radflux} show the
averaged (top) and instantaneous (bottom panel) radiation field in the
radiative simulation. Colors denote magnitude of the radiative flux
and streamlines show its direction. Deep inside the disk photons are
trapped and advected on the BH \citep{sadowski+3d}. At intermediate
angles radiation diffuses out of the disk and contributes to the
radiation in the funnel. The efficiency of accretion in radiation
emitted into the optically thin funnel is not large -- radiation
carries only $\sim 1\%$ of $\dot Mc^2$, what is $\sim 5$ times less
than it would carry for a thin disk accreting on a non-rotating
BH. The corresponding bottom panel shows instantaneous properties of
the radiation field. Radiation escaping through the funnel is quite
laminar, in contrast to turbulent disk interior, where photons are
dragged with optically thick, turbulent gas, producing, on average, the pattern described above.

\begin{table}
\centering
\caption{Properties of the analyzed simulations.}
\label{t.black_holes}
\begin{tabular}{l l l l l l}
	\hline
	\hline
	\textbf{Model} & \textbf{BH spin} & $\mathbf{M}_\mathrm{\mathbf{BH}}$ & \textbf{Radiative} & \textbf{Accretion rate} \\
	\hline
	\multirow{1}{*}{\texttt{hd300a0}} & $a_*=0$ & $10M_\odot$ & No & $\lesssim 10^{-4}\dot{M}_\mathrm{Edd}$  \\
	\multirow{1}{*}{\texttt{d300a0}} & $a_*=0$  & $10M_\odot$ & Yes & $\sim 10\dot{M}_\mathrm{Edd}$ \\
	\hline
	\hline
\end{tabular}
\end{table}

\subsection{Average forces}
\label{s.average}

We begin by calculating the average forces acting on gas at a given location. We averaged the simulations' output on the go, i.e., accounting for states after every single step of time iteration. Each product appearing in Equations~\ref{e.force_thermal}-\ref{e.force_enthalpy} was averaged independently, e.g., 
\be
\left\langle\frac{T^i_\nu-\rho u^iu_\nu}{w}\right\rangle = \frac{\avg{T^i_\nu}-\avg{\rho u^iu_\nu}}{\avg{w}}.
\ee
In such a way we are able to obtain the \textit{average forces} acting at a given location, i.e., the forces that observer sitting at fixed coordinates would see acting on gas flowing by. There is no guarantee that every single parcel of gas will feel exactly these forces when following its own trajectory. However, one can hope that forces obtained in this way will give qualitatively right picture.

\begin{figure}
\centering
	\includegraphics[width=.85\columnwidth]{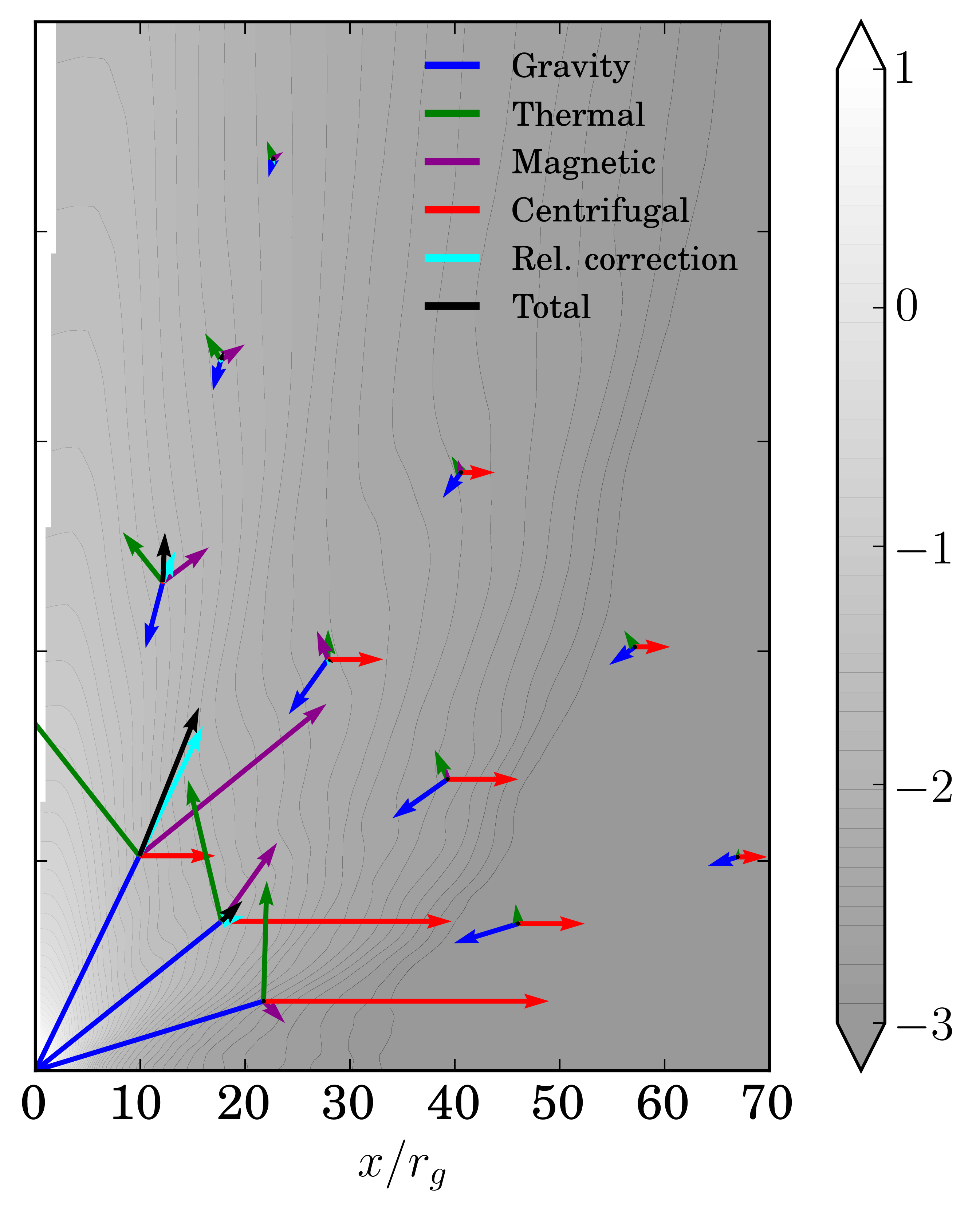}
		\caption{Averaged data from the non-radiative
                  simulation. The figure shows the ratio $\log
                  (b^2/\rho)$ (shades of grey) and the vectorial force distribution at arbitrary points where the arrows indicate both direction and magnitude of the forces.}
	\label{f.hd0300.all_forces.average}
\end{figure}

\begin{figure}
\centering
	\includegraphics[width=.85\columnwidth]{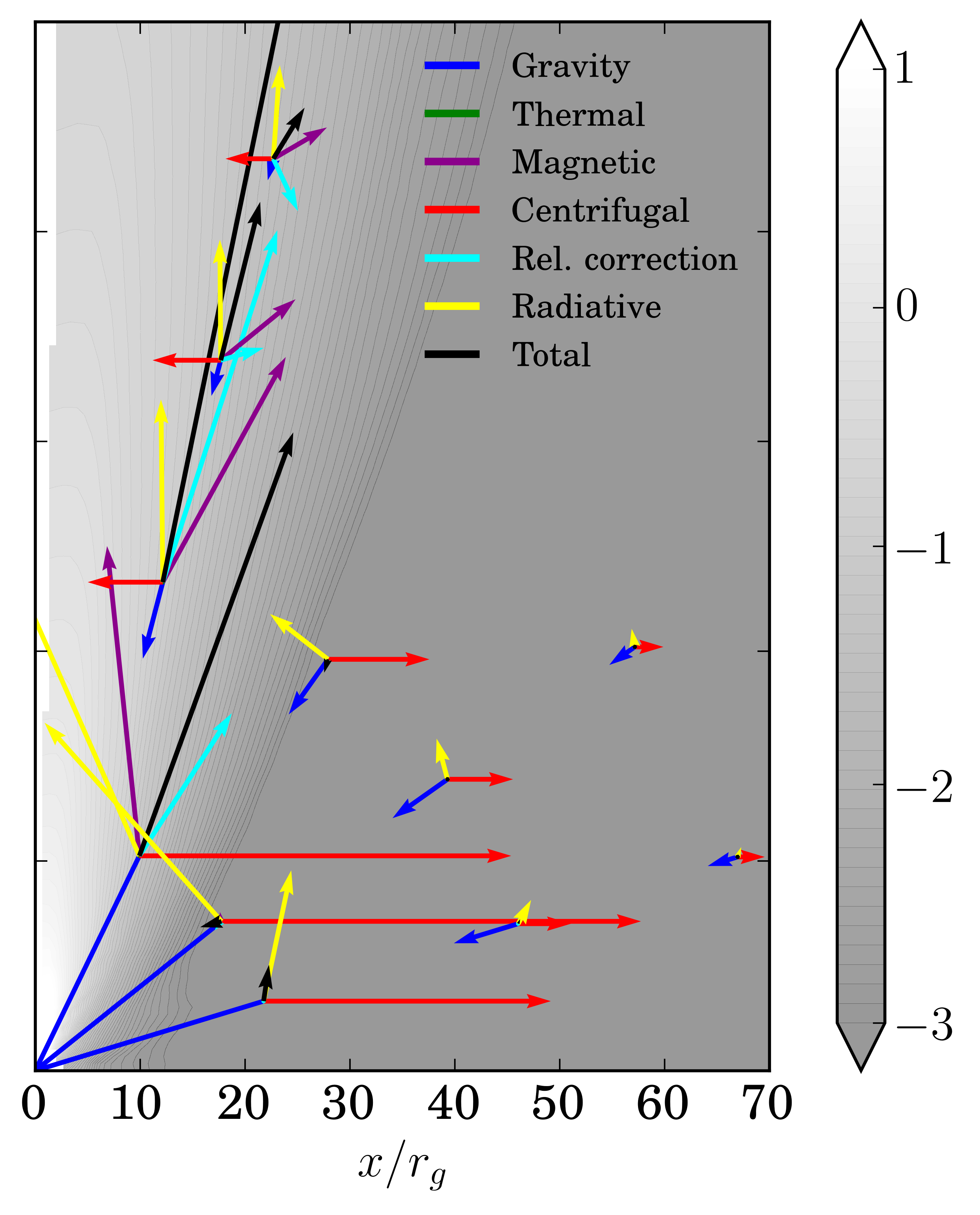}
		\caption{Similar to Figure
                  \ref{f.hd0300.all_forces.average} but for the averaged data from the radiative simulation.}
	\label{f.d0300.all_forces.average}
\end{figure}

The results of this analysis for the optically thin (\texttt{hd300a0})
and thick (\texttt{d300a0}) simulations are shown in
Figures~\ref{f.hd0300.all_forces.average} and
\ref{f.d0300.all_forces.average}, respectively. The background color
denotes the logarithm of the magnetic to rest-mass energy densities
ratio. This quantity measures how strong the magnetic field is, and in
particular, how much the relativistic enthalpy $w=\rho+p_{\rm
  g}+u_{\rm g}+b^2$ departs from the non-relativistic value $w=\rho$.

Using Equations~(\ref{e.force_thermal})-(\ref{e.force_enthalpy}) we
calculated radial and polar components of each force and transformed
them to the orthonormal frame. Acceleration induced by each force
component are plotted with vectors. Blue arrows correspond to the
gravitational force (Eq.~\ref{e.force_gravity}), green to the thermal
pressure gradient force (Eq.~\ref{e.force_thermal}), purple to the magnetic
forces (Eq.~\ref{e.force_magentic}), red to the centrifugal force
(Eq.~\ref{e.force_centrifugal}), and cyan to the relativistic
correction (Eq.~\ref{e.force_correction}). In the case of the
optically thick, radiative run
(Fig.~\ref{f.d0300.all_forces.average}), an additional set of yellow
arrows denote the radiative force (Eq.~\ref{e.force_radiation}). In
both plots the black arrows correspond to the sum of all the forces,
i.e., to the total acceleration acting on gas. The larger is the
length of this vector, the faster gas particles gain velocity. All the
vectors are scaled in the same way, with the scaling factor chosen
arbitrarily to fit the arrows in the panels.

\subsubsection{Optically thin disk}

Figure \ref{f.hd0300.all_forces.average} shows the average forces in the optically thin, purely magnetohydrodynamical simulation. The two forces which are most obvious are the gravitational pull (blue) and the centrifugal force (red arrows). Gravitational attraction always works towards the BH and its magnitude decreases (up to relativistic correction) with the square of radius. The centrifugal force acts only in the direction perpendicular to the axis of rotation. Its strength depends on the value of azimuthal velocity and the distance from the axis. The largest centrifugal acceleration acts on particles close to the BH and located near the equatorial plane, inside the almost Keplerian disk. It is much less significant in case of slowly rotating gas in the polar region.

Outside of the polar region gas moves in a turbulent way with
relatively low velocities and the velocity changes it undergoes are
much less violent then for a free-falling particle. It is therefore
reasonable to expect that the net force acting on the gas in the bulk
of disk would be much smaller in magnitude than the gravitational and
centrifugal forces. To satisfy this condition one has to provide
another force, or forces, which will balance the two. In case of
optically thin disk it is the thermal pressure force (green vectors)
which acts this way. It provides the necessary vertical component
which balances the vertical component of gravitational pull. As a
result, the total acceleration (black arrows) is small in the bulk of
the disk.

Situation is different in the polar region where the gas is
accelerated rapidly and reaches significant radial velocities. In this
case, the magnitude of the net acceleration may be even comparable
with the gravitational pull. In contrast to particles inside the disk, the
centrifugal force is no longer substantial. Instead, the magnetic
force (purple arrows) affects significantly the acceleration. Together
with the thermal force, it provides the vertical force that balances
and overcomes the gravitational pull. For the gas in the innermost
polar region, where the magnetic energy density becomes comparable
with the rest-mass energy density, one should not neglect the
relativistic corrections (cyan arrows), which may reach magnitudes
comparable with the other forces.

\subsubsection{Optically thick disk}

Figure \ref{f.d0300.all_forces.average} shows average forces acting on gas in the optically thick simulation. 
The colors once again denote the strength of the magnetic field. Because of much lower density of gas in the polar region of this simulation than in the case of optically thin disk, the magnetic field now dominates over rest-mass energy and satisfies $b^2/\rho > 1$ in most of the funnel region.

The properties of the gravitational pull (blue arrows) and the centrifugal force (red arrows) are quite similar to the described above - the former points always towards the BH, while the latter is perpendicular to the axis. The centrifugal force points outward everywhere but for the points located in the funnel region. This surprising effect comes from the fact that the $T^\phi_\phi$ component of equation (\ref{e.force_centrifugal}) changes sign in that region because of predominant magnetic field component $b^\phi b_\phi$ (if we had defined this force as proportional to $\rho u^\phi u_\phi$, instead of $T^\phi_\phi$, the force would always point outward, as in the non-relativistic limit). Similar effect was not present in the optically thin case because magnetic field in the polar region was much weaker than in the case described here.

Relatively slow gas velocities in the bulk of the disk results from weak acceleration. This is once again the reason why the forces deep in the disk balance each other and the resulting average net force (and acceleration) is very weak (in most cases the corresponding arrow is not visible in the plot). However, on the contrary to the optically thin case, it is not the thermal pressure gradient which balances the vertical component of the gravitational force. In case of optically thick, radiation-pressure dominated disk like the one simulated, it is the radiation pressure force (equation~\ref{e.force_radiation}, yellow vectors in Figure \ref{f.d0300.all_forces.average}) which provides necessary vertical force. In the optically thick disk it follows the gradient of radiative energy density and therefore points away of the equatorial plane towards the axis. In other words, the radiative force has replaced the thermal pressure force and is now supporting the disk against gravity.

The properties of the forces change significantly in the polar region. As shown in the left panels of Fig.~\ref{f.snapshots.velocity.radflux}, strong flux of radiation coming from the innermost region is streaming out of the system along the axis. Gas which enters this region is immediately swept up and with mildly-relativistic velocities moves away from the BH through the funnel. This radiative acceleration is clearly seen in  Fig.~\ref{f.d0300.all_forces.average} -- radiative force in the polar region has  significantly larger magnitude than the gravitational force at a given location and always points away from the BH. However, the radiative force is not the only one pushing the gas away from the BH. The magnetic force (equation \ref{e.force_magentic}) and the relativistic correction (equation \ref{e.force_enthalpy}) also point upward and reflect the magnetic acceleration of gas in the magnetic tower that develops when magnetic field lines brought into the innermost region are wound up by disk rotation and expand vertically in the polar region dragging the gas behind. This dragging effect, and the resulting magnetic acceleration, is more effective in the optically thick case than it was in the other simulation because of lower density of the gas and larger magnetic to rest-mass energy ratio.

\subsection{Trajectory approach}
\label{s.trajectory}

The averaged data that we based on in the previous section reflects average forces acting at a single \textit{point} over the course of the simulation, and does not reflect forces affecting a particular moving \textit{particle} that happens to pass by the point in question at the specific time. What is more, not all the gas crossing given location, that on average belongs to the outflow region, is guaranteed to escape to infinity. There is also gas which either moves (probably temporarily) inward or is not energetic enough (is bound) to escape the BH gravitational pull. It is therefore not straightforward to extend the analysis done in the previous section, which was based on averaged disk properties, to every single gas parcel. The force balance may be significantly different for outflowing and inflowing gas crossing the same location.

We therefore traced a series of particle trajectories for each
simulation. We chose the trajectories to represent the jet, wind and
disk region of the outflow. We injected
virtual test particles and tracked their trajectories by interpolating
the velocity from snapshot data at time $t$, and thus obtaining the
positions of particles at time $t+\Delta t$. This method holds for
sufficiently small $\Delta t$ intervals. We used snapshot data saved
every $\Delta t =1 GM/c^3$ what gave accuracy good enough to track
even the fastest gas moving with mildly-relativistic velocities $v\sim
0.3c$.  For each trajectory, we then calculated the decomposed
real-time forces along that trajectory, which then were used to determine
the acceleration mechanisms of the outflow

\subsubsection{Optically thin disk}
\label{s.traj.thin}

\begin{figure}
\centering
	\includegraphics[width=.99\columnwidth]{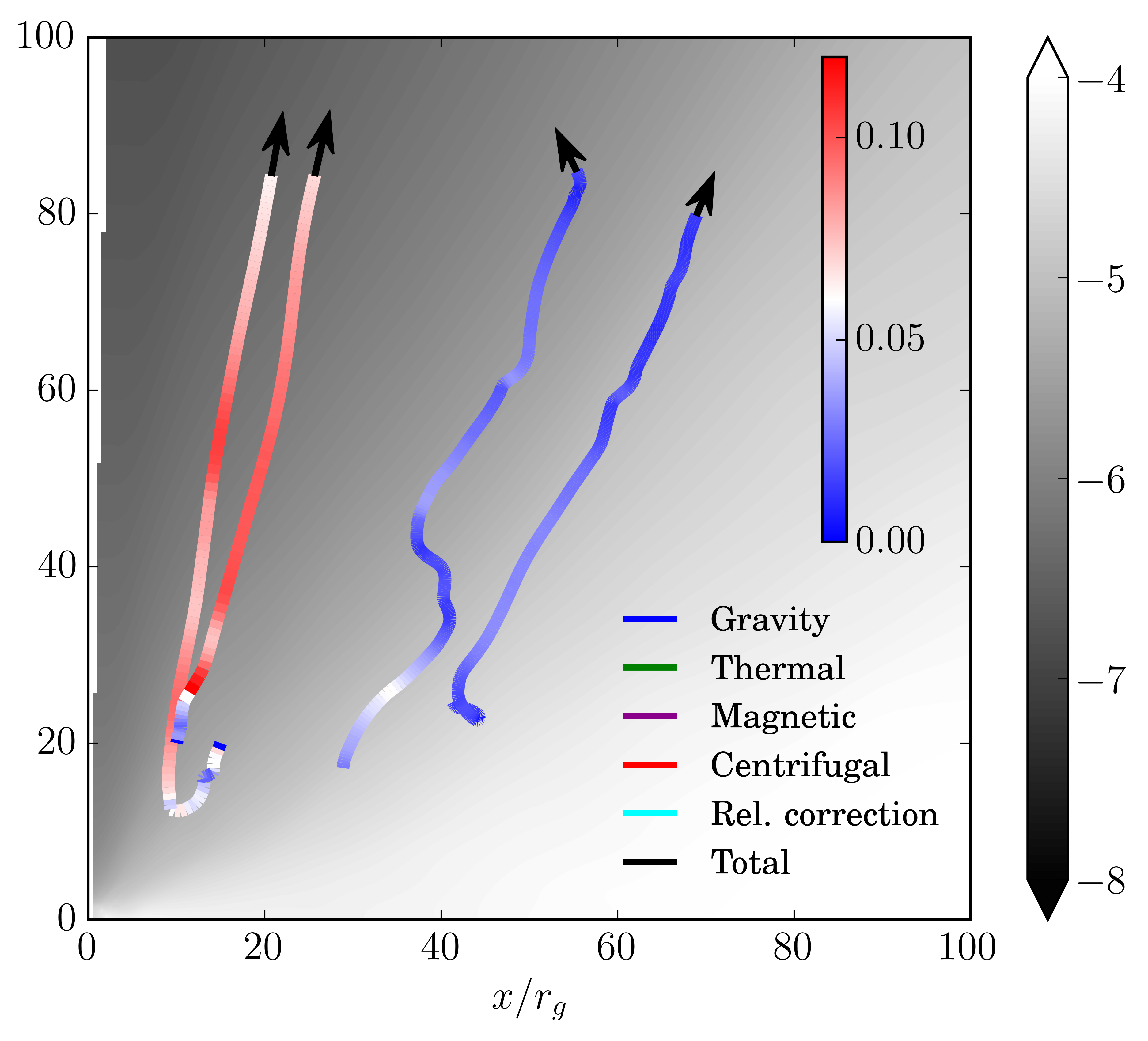}
	\includegraphics[width=.49\columnwidth]{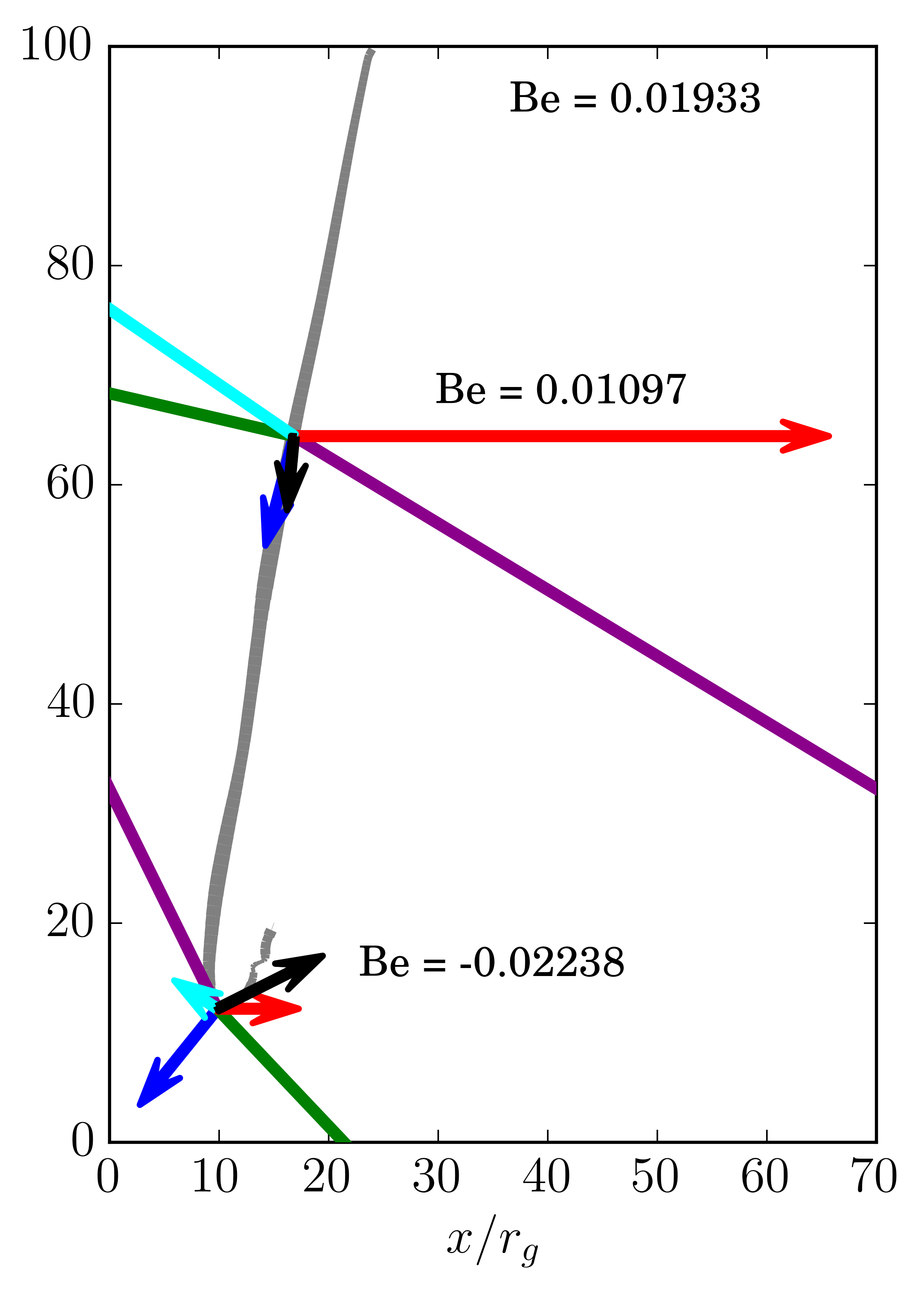}
	\includegraphics[width=.49\columnwidth]{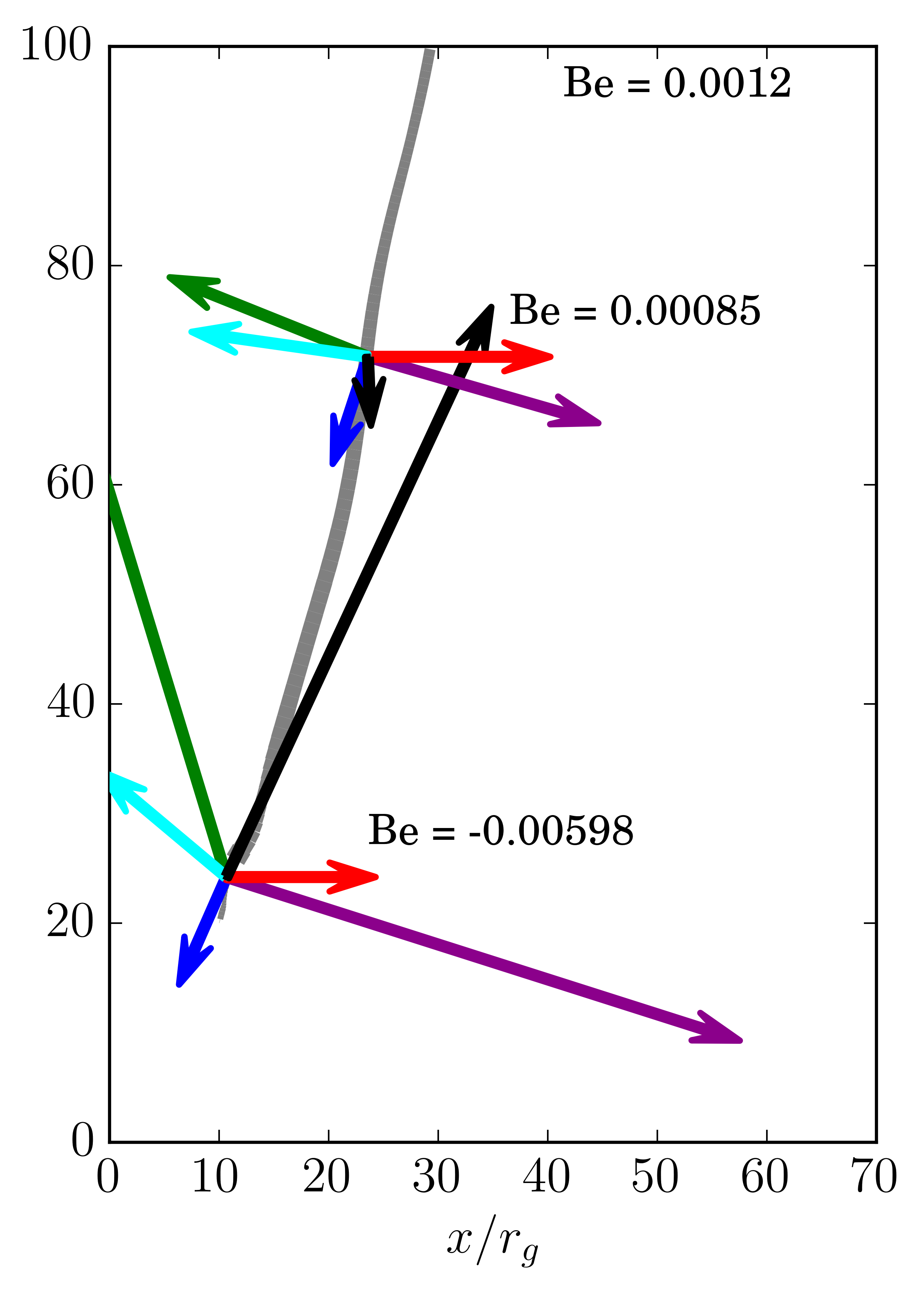}
	\includegraphics[width=.49\columnwidth]{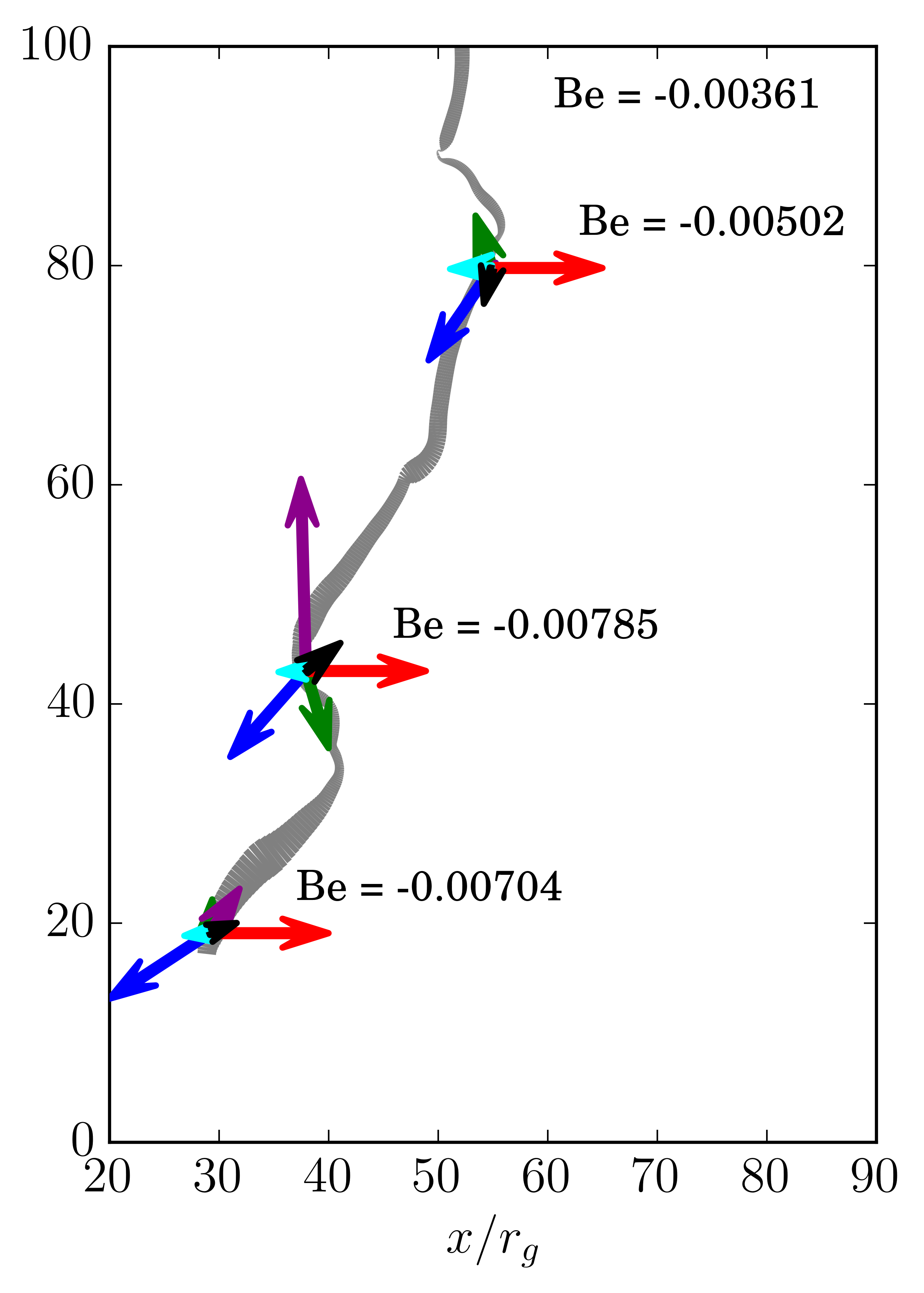}	
	\includegraphics[width=.49\columnwidth]{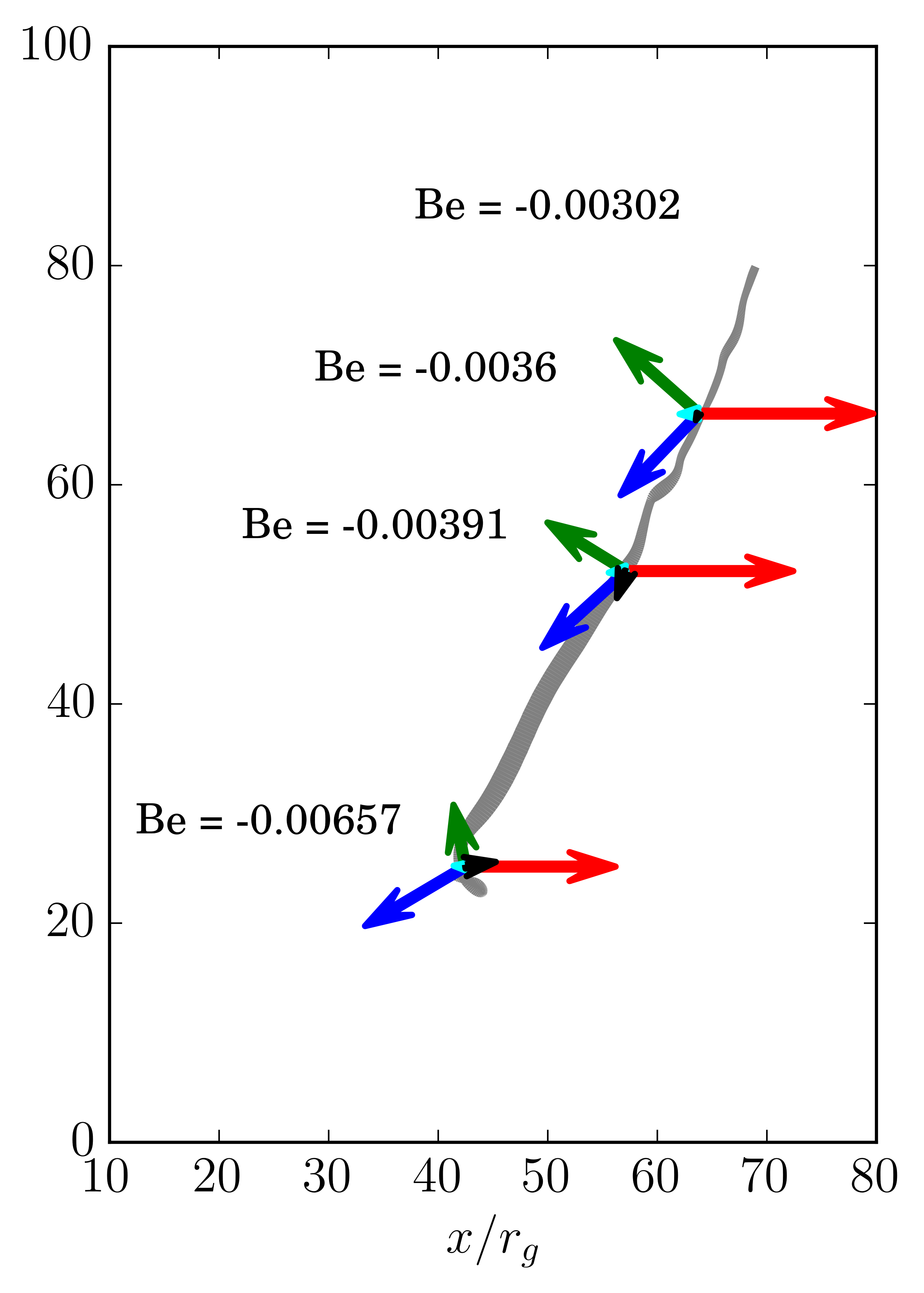}
	\caption{Trajectories for optically thin model \texttt{hd300a0}. The top panel shows two jet and two wind/disk particle trajectories (H1-H4, left to right), where the color is proportional to the speed of the particle in fractions of the speed of light plotted over the logarithm of gas density. The four bottom panels show the force decomposition at arbitrarily chosen points for each individual trajectory. The length of the force vectors is proportional to the force magnitude, scaled by $r^2$, and the width of the trajectory is proportional to the particle velocity (for exact value, see top panel).  Bernoulli number, $Be$, along the trajectories is shown.}
	\label{f.trajectories_hd300a0}
\end{figure}

The top panel in Figure~\ref{f.trajectories_hd300a0} shows the four representative trajectories that we chose for the optically thin model. We will refer to these trajectories as H1 to H4 (numbered left to right). First two (H1 and H2) correspond to gas entering the funnel and escaping along the axis. The other two (H3 and H4) reflect typical trajectories of gas escaping in the wind region of the simulation. The colors along the trajectories reflect the velocity of the gas. As expected, gas in the funnel undergoes the strongest acceleration and reaches velocities of the order of $0.1c$. The wind trajectories (H3 and H4) reflect much slower gas, moving on average with velocities $\sim 0.01c$.

The four panels below the top one show the forces acting on gas along
each of the trajectories. The arrows have the same meaning as in the
previous plots. The numbers on the plots reflect the Bernoulli number
of the gas (equation~\ref{eq:Be}) at each location for which we
calculated the forces. Positive values correspond to unbound gas.

The jet trajectories (H1 and H2)  both accelerate and then decelerate while maintaining a growing Bernoulli number making them part of the real outflow. However, the main initial acceleration mechanism of these two trajectories differ, as trajectory H1 is initially accelerated by the magnetic force (magenta arrows) and trajectory H2 is initially accelerated by the thermal pressure force (green arrows). It is evident that for these trajectories the acceleration is not a continuous process -- gas is rapidly accelerated in the innermost region, it enters the funnel with large velocity, but then the further acceleration ceases and the gravitational pull slows down the gas. There is no unique force which is responsible for this initial acceleration -- it could be either magnetic or thermal force.

Wind trajectories H3 and H4 both exhibit relatively low positive radial velocities and show negative but growing Bernoulli numbers. They are part of the turbulent outflow within our simulation boundaries, but the growing number indicates that they could be part of the real outflow. We do not observe a sudden acceleration, but rather a series of minor acceleration events giving the trajectories on average positive radial velocities.  The main acceleration mechanisms for the trajectory closest to the funnel region (H3) and the one deeper in the disk (H4) are the magnetic and thermal forces, respectively, with the centrifugal force also providing significant component along the trajectory.

\subsubsection{Optically thick disk}
\label{s.traj.thick}

\begin{figure}
\centering
	\includegraphics[width=.99\columnwidth]{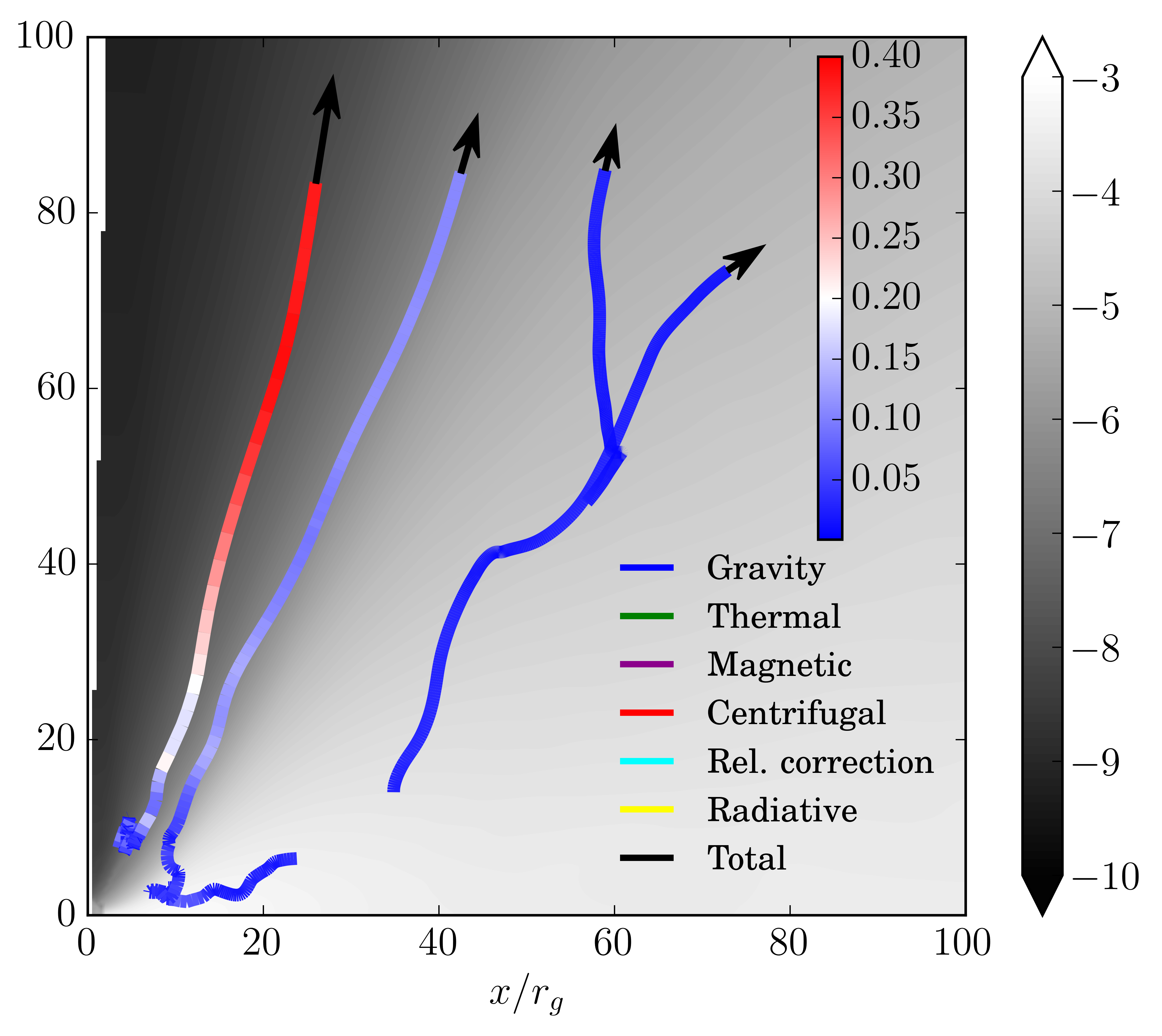}
	\includegraphics[width=.49\columnwidth]{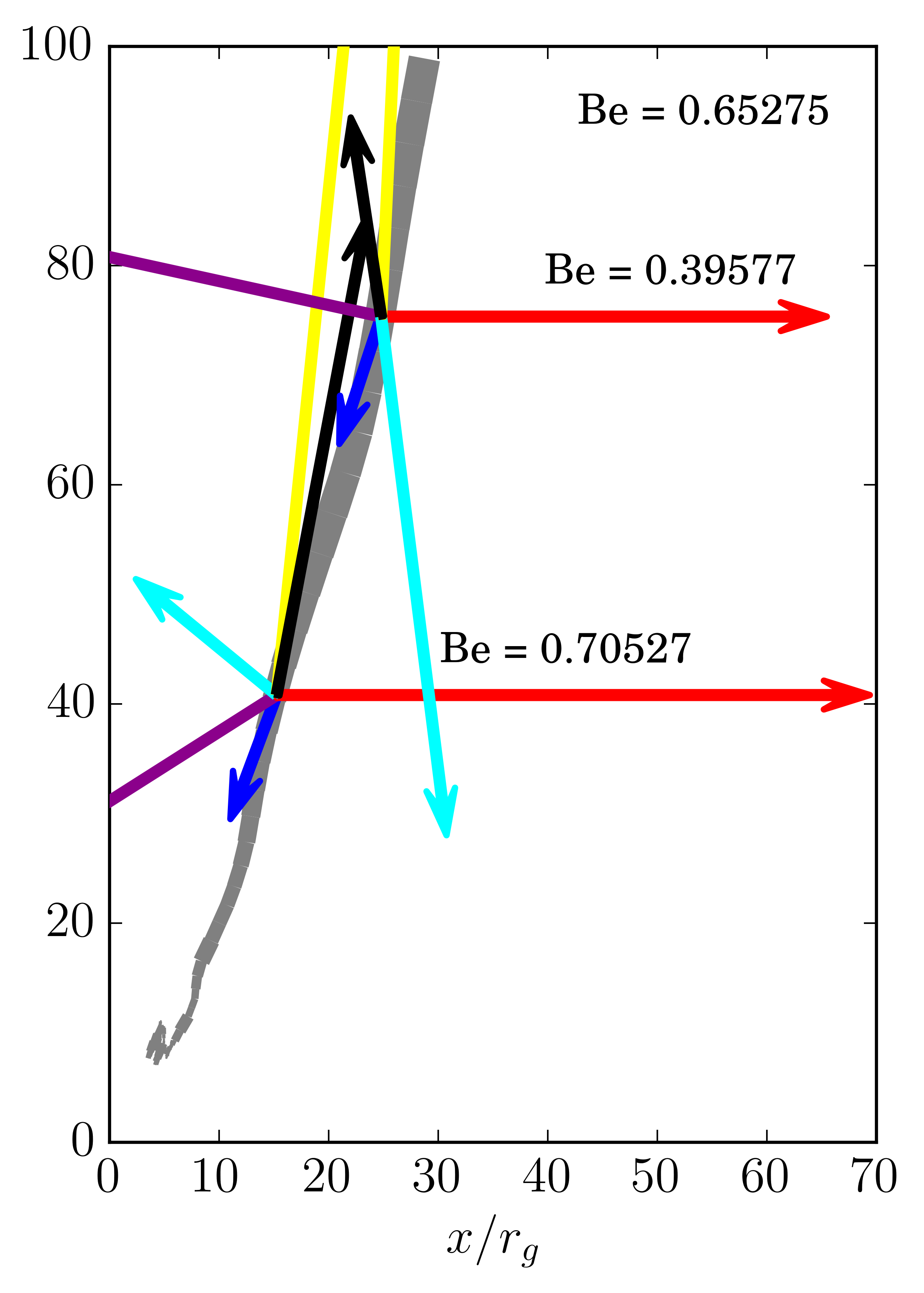}
	\includegraphics[width=.49\columnwidth]{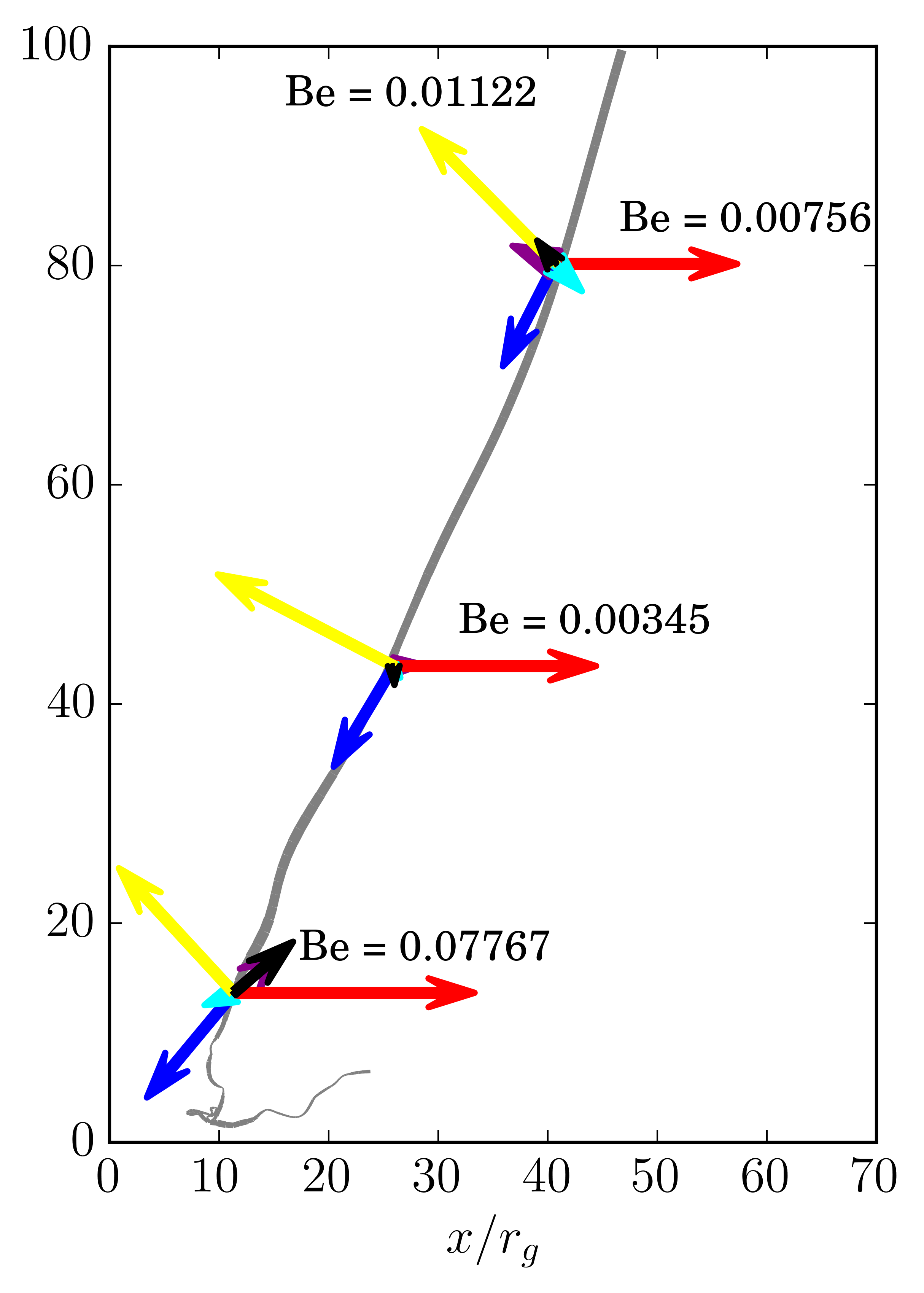}	
	\includegraphics[width=.49\columnwidth]{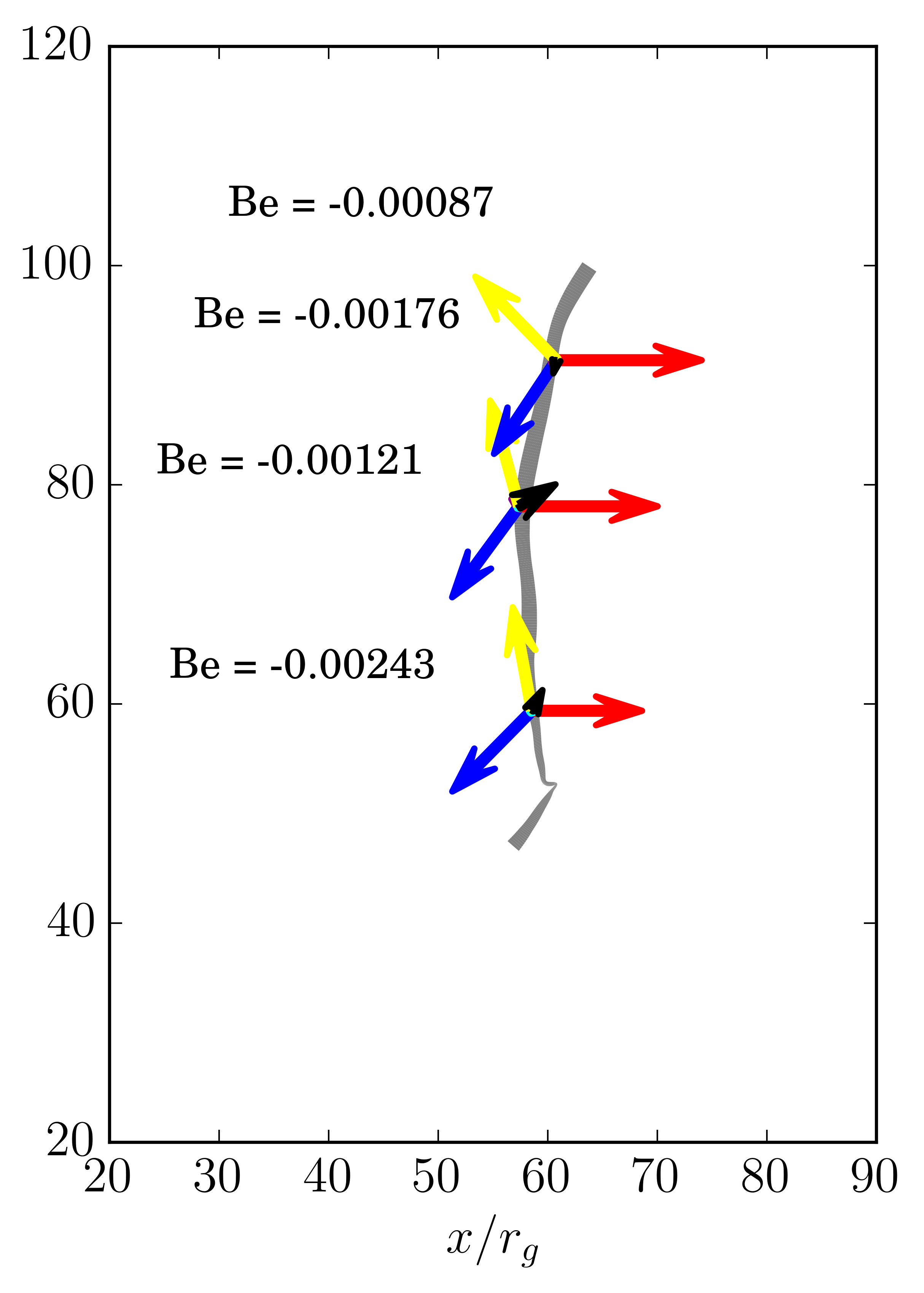}	
	\includegraphics[width=.49\columnwidth]{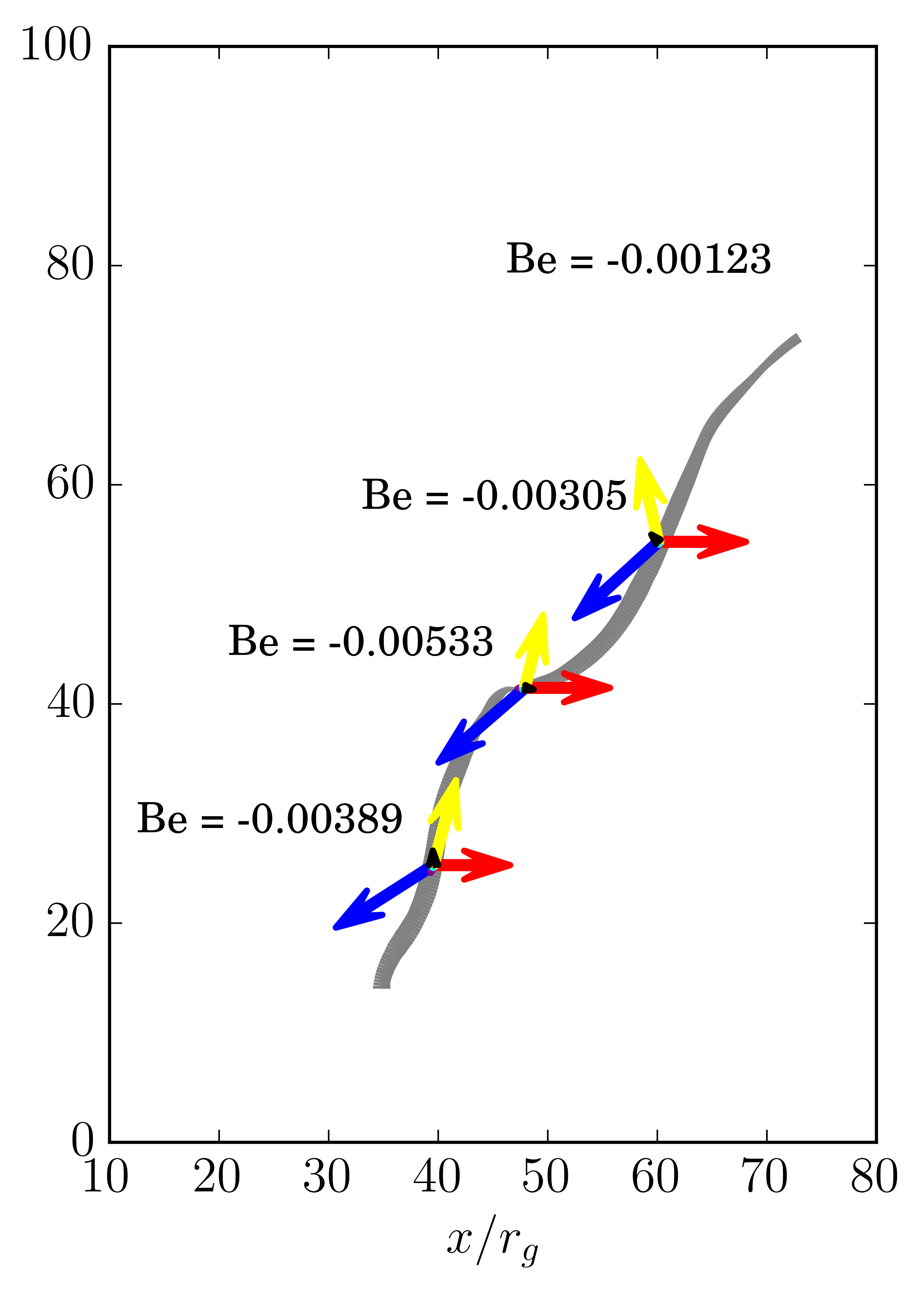}
	\caption{Similar to Figure \ref{f.trajectories_hd300a0} but for the optically thick model \texttt{d300a0} and trajectories R1-R4 (left to right). The additional radiative force is denoted by yellow arrows.}
	\label{f.trajectories_d300a0}
\end{figure}

Figure~\ref{f.trajectories_d300a0} presents in a similar way four outflowing trajectories in the optically thick simulation. We will refer to them as trajectories R1 (leftmost) to R4 (rightmost). The top panel shows the trajectories on top of the average gas density distribution. The colors along them denote the gas velocity. Trajectory R1 stands out as the one closest to the axis -- the gas following this trajectory reaches the highest velocity, $v\sim 0.4c$. The second leftmost trajectory, R2, follows roughly the boundary between the polar funnel and the disk. The corresponding gas is not accelerated so efficiently but still reaches velocity of the order of $\sim 0.1c$. The other two trajectories are deep inside the optically thick region and are much slower with gas not exceeding $0.02c$.

The four panels below the top one show forces acting along each of the trajectories. Gas which managed to enter the funnel region (reflected in path R1) has very low density and is strongly accelerated by the radiation force (yellow arrows) which is responsible for effective and continuous acceleration. This acceleration results in a mildly-relativistic outflow of unbound gas along the axis. This outflow is characterized by lowest densities but may carry significant amount of kinetic energy \citep{sadowski+radjets}.

The gas following the boundary of the funnel and the disk (trajectory R2) is also pushed forward predominantly by the radiation force. This time, however, the force is weaker and is pointing more towards the axis, not along it. This results from the fact that at this location the radiation diffuses out of optically thick disk following the gradient of radiative energy density. Radiative flux, together with the centrifugal force, is strong enough, however, to accelerate the gas and make it unbound.

The wind trajectories (R3 and R4) correspond to gas flowing through surface layers of the disk. The radiative flux in the comoving frame of the gas, which results from photon diffusion, points towards the disk surface and so does the radiative force. The radiative and centrifugal forces are not strong enough, however, to  efficiently accelerate the gas -- it is just enough to move the gas slowly outward. The corresponding Bernoulli numbers are negative but growing and it is possible that ultimately the radiation will manage to make the gas unbound. The properties of the force balance along these two trajectories very well agree with the results of the analysis based on averaged data. 

\section{Discussion}\label{s.discussion}

The outflow in the optically thin, non-radiative simulation
is mainly driven by the magnetic and thermal forces. The initial acceleration mechanism for the radiative simulation is the
radiative force. In highly magnetized and the innermost regions in
both models the relativistic correction
(Eq.~\ref{e.force_correction}) is not negligible, and the standard
Newtonian decomposition of forces would not be adequate.

The jet  trajectories (H1 and H2, see Section~\ref{s.traj.thin}) left the simulation boundary decelerating (gravitational pull was prevailing) but with positive and still growing Bernoulli numbers, making them part of the real outflow, while the wind trajectories (H3 and H4) of the non-radiative simulation were characterized by negative but still growing Bernoulli number. The maximal velocities in the non-radiative model were approximately $0.10 c$. 
The radiative accretion flow  revealed much larger velocities,
approaching $0.40c$ for low-density gas in the jet region accelerated
vertically by the uniform radiation flux. Both
radiative jet trajectories we studied (R1 and R2, see Section~\ref{s.traj.thick}) left the simulation bounds with positive
and growing Bernoulli parameters. As in the non-radiative model, the
gas contained in the wind was bound inside the domain, but the outward
forces were still increasing its Bernoulli number.

In case of the optically thin, non-radiative disks, the initial acceleration seems to play the major part in generating real outflow, as the gas is accelerated rapidly in the inner region and then propagates with forces near equilibrium. The trajectories in the jet region of the radiative simulation, on the contrary, are continuously exposed to radiation force and thus undergo continuous acceleration.

The average forces acting on the gas at a given location reflect
dominating forces acting on a gas parcel along its
trajectory. However, 
the average forces do not allow to catch the exact timing and location
of the short-lived but significant acceleration episodes. They also do
not reflect the turbulent nature of the flow.
The trajectory approach provides full information about the
acceleration mechanism for traced trajectories, making it ideal for
studying this phenomena. While it is not possible to trace every
single gas trajectory, the averaged data serves as a reference of what
one should expect given a large enough set of traced
trajectories.

We have compared an optically thin disk, corresponding to the
radiatively inefficient regime and to the lowest accretion rates, with
a model of optically thick disk accreting at a super-ciritcal rate of $10\Medd$. To get full understanding of processes driving outflows in accretion disks, one would have to extend our work with a similar study covering wider range of accretion rates, including geometrically thin, sub-Eddington disks, and taking rotating BHs into account. Especially in the case of spinning BHs one may expect that the extra energy extracted from the BH spin will significantly alter the acceleration mechanisms in the jet region.

Similar modelling of forces acting on gas in accretion flows was done previously in~\cite{yuan2015numerical} and \cite{takahashi2015radiation}.
The former work focused solely on optically thin disks, while the
latter analyzed only a super-critical disk similar to our radiative model. Our conclusions are in general agreement with both these works. However, we calculated and decomposed the forces in fully relativistic way and directly compared driving mechanisms in optically thin and thick disks.

\section{Summary}

We have analyzed the force balance and the resulting acceleration of the outflow in two simulations of a BH accretion flow. Both were performed in general relativity with a state-of-the-art numerical methods and assumed zero BH spin and weak, non-saturated magnetic field. One model corresponded to radiatively inefficient, optically thin disk, while the other to a super-critical disk with accretion rate of $10 \Medd$. We found that:

\begin{enumerate}

\item \textit{Gas acceleration:} Driving gas out of the disk is not, in most cases, a continuous and simple process. There are usually only a couple of short-lived episodes which result in significant outward acceleration of the gas. But for them, the gas velocity fluctuates, as does the net force acting on gas. An exception is the polar region in radiative disks, where gas is constantly pushed away by radiation pressure.

\item \textit{Outflows in optically thin disks:} Radiatively
  inefficient accretion produces outflows at large range of
  angles. The gas in the polar region is accelerated mostly by the
  magnetic forces and reaches velocities of the order of $0.1c$. At
  larger distance from the axis, it is the thermal pressure and
  centrifugal forces which compensate for gravity and push the gas out
  of the disk. The magnetic forces are not significant but for the
  most polar region. Therefore, the outflow outside of the polar
  region is not driven by the magnetocentrifugal acceleration.

\item \textit{Outflows in optically thick disks:} The fastest outflows in radiative disks take place near the axis, in the optically thin funnel region. Gas can reach there mildly-relativistic velocities exceeding at times $0.4c$. Such significant acceleration results mostly from the radiation pressure force pushing the gas along the axis. Magnetic acceleration is also important in the polar region. In the optically thick disk it is the radiation pressure, not the thermal pressure as in the other case, which balances gravity and pushes gas out of the disks, either into the funnel or radially outward. The magnetic acceleration is once again not significant but for the most polar region.

\item \textit{Relativistic corrections:} We found that the relativistic correction that we have singled out from other forces (Eq.~\ref{e.force_enthalpy}) contributes significantly in the regions close to the BH and where magnetic field energy is comparable or exceeds the rest-mass energy density. Therefore, force decomposition based on non-relativistic formulae would be inadequate in these regions.

\item \textit{Average vs trajectory approach:} Only the study of balance of forces based on following particular gas trajectories is able to catch the non-uniform nature of gas acceleration.

\end{enumerate}

\section{Acknowledgments} 

This work was completed during the annual Research Science Institute (RSI) workshop for high school graduates held at Massachusetts Institute of Technology. AM thanks RSI for the support, and MIT Kavli Institute for Astrophysics and Space Research, as well as Harvard-Smithsonian Center for Astrophysics, for their hospitality.
AS acknowledges support
for this work 
by NASA through Einstein Postdoctoral Fellowship number PF4-150126
awarded by the Chandra X-ray Center, which is operated by the
Smithsonian
Astrophysical Observatory for NASA under contract NAS8-03060.
The authors acknowledge computational support from NSF via XSEDE resources
(grant TG-AST080026N), and
from NASA via the High-End Computing (HEC) Program
through the NASA Advanced Supercomputing (NAS) Division at Ames
Research Center.
 
\nocite{*}

\bibliographystyle{apj}


\end{document}